\documentclass[aps,prd,superscriptaddress,nofootinbib,eqsecnum]{revtex4}

\pdfoutput=1

\usepackage[footnotesize]{caption}
\usepackage{amsmath,amsfonts,amssymb,latexsym}
\usepackage{graphicx,color}
\usepackage{color}
\usepackage{float}
\usepackage{hyperref}
\usepackage{subfigure}
\begin{document}

\title{Identifying Universality in Warm Inflation}

\author{Arjun Berera} \email{ab@ph.ed.ac.uk} 
\affiliation{School of Physics and Astronomy, 
University of Edinburgh, Edinburgh, EH9 3JZ, United Kingdom}

\author{Joel Mabillard} \email{joel.mabillard@ed.ac.uk} 
\affiliation{School of Physics and Astronomy, 
University of Edinburgh, Edinburgh, EH9 3JZ, United Kingdom}

\author{Mauro Pieroni} \email{mauro.pieroni@uam.es}
\affiliation{Instituto de F\'{\i}sica Te\'orica UAM/CSIC C/ Nicol\'as Cabrera 13-15 Universidad
Aut\'onoma de Madrid Cantoblanco, Madrid 28049, Spain}

\author{Rudnei O. Ramos} \email{rudnei@uerj.br}
\affiliation{Departamento de F\'{\i}sica Te\'orica, Universidade do
Estado do Rio de Janeiro, 20550-013 Rio de Janeiro, Rio de Janeiro, Brazil}

\begin{abstract}

Ideas borrowed from renormalization group are applied
to warm inflation to characterize the inflationary
epoch in terms of flows away from the de Sitter regime.
In this framework different models of inflation fall into universality
classes.  Furthermore, for warm inflation this approach also helps to
characterise  when inflation can smoothly end into the radiation
dominated regime. Warm inflation has a second functional dependence compared to
cold inflation due to dissipation, yet despite this feature, it
is shown that the universality classes defined for cold inflation
can be consistently extended to warm inflation. 

\end{abstract}

\maketitle

\tableofcontents
\section{Introduction}
\label{sec:introduction}
Observational cosmology, as demonstrated by the recent Planck results,
has reached an impressive  level of precision that can
set constrains on many cosmological models, including
inflation~\cite{Planck2015,Ade:2015ava,Ade:2015xua}.  However, despite the level
of accuracy achieved by Planck, the degeneracy problem of
inflationary model building still persists, in that
many inflationary models can produce predictions, like for the
tensor-to-scalar ratio and for the spectral tilt, that are very similar
and compatible with the data. In Ref.~\cite{Binetruy:2014zya} the idea of
universality classes was suggested as a means to classify a wide range
of inflation models, and thus subsumes a large number
of them in terms of their salient properties relevant to observation.
This approach borrows ideas from the renormalization group (RG) methods of quantum field theory (QFT), such as the concept of flow away from a fixed point, here corresponding to the exact de Sitter (dS) geometry, and the use of an analog to the renormalization group equation (RGE) for the $\beta$-function. 

The $\beta$-function formalism was introduced in
Ref.~\cite{Binetruy:2014zya} and further developed and extended in Refs.~\cite{Pieroni:2015cma,Pieroni:2016gdg,Binetruy:2016hna,Cicciarella:2016dnv,Cicciarella:2017nls},
to identify universality among the wide zoology of inflationary
models. This formalism is based on the application of the
Hamilton-Jacobi (HJ) approach to cosmology~\cite{Salopek:1990jq}. It
relies on a formal analogy 
between the equation describing the
evolution of a scalar field in an expanding background and a
RGE of QFT. As will
be explained below, this analogy is not coincidental but has 
underpinnings with holography.
In this framework the near scale invariance experienced by the Universe
during inflation is interpreted as a departure  of the
corresponding RGE from a fixed point corresponding to an exact dS spacetime. A single parametrization of the $\beta$-function, close to the dS fixed point, thus defines an universality class of
models that can be grouped together, sharing a single asymptotic
behavior. As a consequence, arbitrary potentials can be classified
into a small set of classes according to the behavior of their
associated $\beta$-function in the neighborhood of the fixed point.

This approach has some direct advantages. {}First of all, by grouping
different potentials into a small set of classes, it significantly
reduces  the number of relevant cases to consider. {}Furthermore, as
the formalism relies on intrinsic properties of inflation, it is
completely general and in particular, it does not assume
slow-roll, since for example it has been successfully applied to
constant-roll inflation~\cite{Cicciarella:2017nls}.
{}Finally, as mentioned already,
this formalism has deep theoretical
motivations arising from the holographic description of the Early
Universe (see, for example,
Refs.~\cite{Skenderis:2006jq,McFadden:2009fg,McFadden:2010na}). Within
the (A)dS-CFT correspondence of Maldacena~\cite{Maldacena:1997re}, the
flow away from the dS fixed point, which is realized during inflation,
is dual to a deformation of the associated conformal field theory (CFT) due to relevant or
marginal operators. By applying these methods to describe the early
Universe, and in particular inflation, it is both possible to shed
a new light on some of its problematic aspects and to
provide an alternative interpretation of the observational
constraints~\cite{McFadden:2010vh,Bzowski:2012ih,Garriga:2014ema,Garriga:2014fda,Afshordi:2016dvb,Afshordi:2017ihr,Hawking:2017wrd,Conti:2017pqc}.

The aim of the present work is to formulate warm inflation in terms of
the $\beta$-function formalism.  Warm
inflation~\cite{Berera:1995wh, Berera:1995ie} differs from the
usual paradigm of cold inflation in the fact that dissipative processes can lead to a
sustainable radiation production throughout the inflationary
expansion. Warm inflation will happen for regimes of parameters  such
that the inflaton interactions with other field degrees of freedom are
not negligible and they generate dissipation terms, such that a  small
fraction of vacuum energy density can be converted to radiation. When
the magnitude of these dissipation terms are strong enough to
compensate the redshift of the radiation by the expansion, a
steady state can be produced, with the inflationary phase happening in
a thermalized radiation bath. There have been many constructions based
on particle physics models demonstrating the viability of
this special regime of inflation, see, for example,
Refs.~\cite{Berera:1998px,Berera:2002sp,Bastero-Gil:2016qru} and for a
review, see also Ref.~\cite{Berera:2008ar}.  
Recently a first principles warm inflation model was constructed from
QFT which involves just a few fields~\cite{Bastero-Gil:2016qru}, thus convincingly demonstrating that
warm inflation models are on an equal footing to cold inflation as model building prospects.  Moreover, the
dissipative effects and the presence of a non-vanishing radiation bath
are able to change both the inflationary dynamics at the background
and at the fluctuation
levels~\cite{Ramos:2013nsa,Bartrum:2013fia,Bastero-Gil:2014jsa,Bastero-Gil:2014raa,Moss:2007cv,Vicente:2015hga, Arya:2017zlb, Bastero-Gil:2017wwl,Rangarajan:2018tte}, such that there can be distinctive differences
between the two paradigms which could be testable.  As such, it
is useful to understand the dynamical structure of warm inflation
through different perspectives, which is a motivation of this paper.

By applying the $\beta$-function formalism to warm inflation, we show
that there are two intervening characteristic functions regulating the
dynamics. One of them is the function already identified in
Ref.~\cite{Binetruy:2014zya}, which was defined in the cold inflation
case, and which controls the way  the inflaton drives the departure from the dS fixed point. In the warm inflation context, we show that another
function controlling the level of radiation production naturally emerges. 
By following the evolution of these two
functions, we are able  not only to fully characterize  the dynamics,
but also to determine when the end of warm inflation smoothly
connect with the radiation dominated regime.  Furthermore, these two
functions allow us to group different forms of inflationary potentials
in certain universality classes.  Since this description sets direct control on the dynamics by using parameters which are different from the
usual slow-roll coefficients, it offers an extremely powerful method to
describe the inflationary evolution (and its end) in an independent
and novel way.

In this work we make use of the generalized framework offered by the $\beta$-function formalism to obtain an analytical understanding of warm inflation. We first show that in some toy models a full analytical description of warm inflation can be derived and then we focus on more realistic scenarios. In particular, we show that it is possible to derive a relatively accurate description of both the weak and strong dissipative regimes. Among the main results of the paper there is the observation that, despite a second functional dependence is introduced, a universal description of inflation, similar to the one of Ref.~\cite{Binetruy:2014zya}, can still be consistently formulated. This allows then to study the effect of the various forms of the dissipation terms commonly considered in warm inflation on the classes of universality and on their predictions for the scalar spectral index and the tensor-to-scalar ratio. Remarkably, we show that within the $\beta$-function formalism it is easy to identify the degeneracy in the inflationary observables for some models with different dissipation coefficient forms.

This work is organized as follows.  In Sec.~\ref{sec:general_warm} we
give the basic equations describing the dynamics of warm inflation
along also the expression of the scalar curvature power spectrum for some
representative examples of warm inflation dissipative forms.  In
Sec.~\ref{sec:warm_beta} we describe how the $\beta$-function formalism is applied to warm inflation. We provide details on the method and present explicit examples  in Sec.~\ref{sec:methods_and_results}. The results are presented and discussed in Sec.~\ref{sec:discussion_of_results}. Our concluding remarks  and future perspectives are given in Sec.~\ref{sec:conclusions_and_future}.  An Appendix is
included to provide further useful formulae and present some different inflation classes.

\section{Warm inflation basics}
\label{sec:general_warm}
Let us restrict to the case of a single inflaton field with pressure and energy density respectively given by
\begin{equation}
	\begin{aligned}
	p_\phi &=\frac{1}{2}\dot{\phi}^2-V(\phi)\;, \\
        \rho_\phi &=\frac{1}{2}\dot{\phi}^2+V(\phi)\; ,
        \end{aligned}
\end{equation}
where $V(\phi)$ is the scalar potential. In this case the dynamics of warm inflation at the background level is governed by
the equation of evolution for the inflaton field
\begin{eqnarray} \label{eq:kleingordon}
	\ddot\phi + (3H+\Upsilon)\dot\phi + V_{,\phi}=0\;,
\end{eqnarray}
where $\Upsilon$ is the dissipation coefficient, which in general can
be a function of both the temperature and the background inflaton
field $\phi$, by the equation for the evolution for the radiation energy density
$\rho_r$
\begin{eqnarray} \label{eq:continuitygauge}
	\dot\rho_r + 4H\rho_r = \Upsilon\dot\phi^2\;,
\end{eqnarray}
and by the Einstein equations (in units of reduced Planck mass $  M_P /\sqrt{8\pi} = m_P = \kappa^{-1} = 1$)
\begin{eqnarray}
3 H^2 & = &  \rho_{\phi} + \rho_{r}\;, \label{eq:friedmann}  \\
-2 \dot{H} & = &  p_{\phi} + \rho_{\phi} +p_r+
	\rho_{r}=\frac{4}{3}\rho_r+\dot{\phi}^2\; \label{eq:raychauduri}.
\end{eqnarray}
Notice that one of these equations is redundant. Moreover, it would be equivalently possible to use the equation of continuity for the inflaton energy density $\rho_\phi$
\begin{equation}
	\dot{\rho}_\phi+3H\dot{\phi}^2=-\Upsilon \dot{\phi}^2\;.
	\label{eq:continuityphi}
\end{equation}
In the following, we will be working mainly with the dissipation coefficient ratio $Q$, defined as $Q\equiv \Upsilon/3H$. Typically, the study of warm inflation assumes the radiation to be thermalized, i.e.
\begin{equation}
	\label{eq:p_phi_rho_phi_rho_T}
	\rho_r=\frac{\pi^2 g_*}{30}T^4=3p_r\;,
\end{equation}
where $g_*$ being the number of relativistic degrees of freedom for the
radiation bath.  Without knowing the details on the
microphysics of the system the condition to attain this situation is normally
$Q \gtrsim 1$. Namely, we require the radiation
fields, generically associated with relativistic degrees
of freedom, e.g., gauge or other fields, to have at
least one interaction mediated by the inflaton  in a Hubble time.
This is typically the necessary condition for thermalization. However, there
could be other mechanisms beyond the radiation fields/inflaton
interaction, which can dramatically modify this setup. {}For
  example, if the radiation fields are coupled to the standard model
  (SM),  the Schwinger process~\cite{Schwinger:1951nm} should provide
  an extremely efficient mechanism to reach thermalization, see for
  example Refs.~\cite{Hayashinaka:2016qqn,Tangarife:2017rgl,Ferreira:2017lnd}. 
  However, in this work we are not interested in building models
where these mechanisms are taken into account and we 
assume that some process ensures that radiation thermalizes. For a
detailed quantification of the thermalization  process relevant for
warm inflation and done in the context of the Botzmann equation, see,
e.g., Ref.~\cite{Graham:2008vu}. Therefore, we assume Eq.~\eqref{eq:p_phi_rho_phi_rho_T} to hold. Note also that the specific form
for the dissipation coefficient $\Upsilon$ in the above equations can
only be determined by the details of the microphysics during
inflation. Different forms of dissipation coefficients derived from
QFT have been derived explicitly  e.g. in
Refs.~\cite{BasteroGil:2010pb,BasteroGil:2012cm}. It is also worth mentioning that warm inflation helps in easing the $\eta$-problem~\cite{Berera:1999ws, Berera:2004vm} since in the strong dissipative regime $Q\gg1$ the inflaton mass is larger than $H$.

\subsection{The scalar spectrum of perturbations in warm inflation}
\label{sec:warm_perturbations}

Given the complexity of the warm inflation dynamics, which involves a
system of coupled fluids associated with the inflaton and radiation,
alongside perturbations, which in general are  also
coupled~\cite{Graham:2009bf,BasteroGil:2011xd}, an analytical treatment
for the spectrum of perturbations is in general difficult. In what follows, we briefly present this analysis. For a full discussion we send the reader to the Refs.~\cite{Ramos:2013nsa, Taylor:2000ze}. The dimensionless scalar
power spectrum $\Delta_s^2(k,\tau)$ at horizon crossing, meaning $k \tau = 1$ where $\tau$ is used to denote the conformal time, is a sum of thermal and vacuum contributions
\begin{equation}
	\Delta_s^2(k,\tau) = \Delta_{s,th}^2(k,\tau) +
        \Delta_{s,vac}^2(k,\tau) \;.
\end{equation}
In the simplest realization of inflation
$\Delta_{s,vac}^2(k,\tau)$ at horizon crossing can be simply expressed
as
\begin{equation}
	\label{eq:vacuum_contribution}
	\left. \Delta_{s,vac}^2(k,\tau) \right|_{\tau = k^{-1}} =
        \frac{1}{4 \pi^2} \frac{H^4}{\dot{\phi}^2} \;.
\end{equation}
On the other hand, the thermal contribution in general is something
that depends on the microphysics of the model.  Nevertheless, it is known that semi-analytical expressions for the full spectrum of scalar perturbations can be derived. In particular the spectrum at horizon crossing can be expressed as~\cite{Berera:1995wh, Berera:1999ws, Benetti:2016jhf} 
\begin{equation}
  \label{eq:modified_scalar}
	\left. \Delta_s^2(k,\tau) \right|_{\tau = k^{-1}} =
        \left. \left( \frac{H^2}{2\pi \dot{\phi}}\right)^2 \left[ 1 +
        2 n_{BE} (T_{\delta \phi}) + \frac{\sqrt{12} \pi {Q} }{\sqrt{3 + 4 \pi {Q} }}
        \frac{T}{H} \right] G(Q) \right|_{\tau = k^{-1}} \; ,
\end{equation}
where $n_{BE} = \left[\exp(H/T_{\delta \phi}) - 1 \right]^{-1}$ is the Bose-Einstein distribution and $G(Q)$ is a function of $Q $ that accounts for the fact that the radiation fluctuations are in general coupled to the inflaton which is thus leading to a growing mode in the inflaton fluctuations~\cite{Graham:2009bf, BasteroGil:2011xd, BasteroGil:2009ec}. Moreover, the temperature $T_{\delta \phi}$ inside $n_{BE} $ corresponds to the temperature of the inflaton fluctuations and is not
necessarily the same as $T$, corresponding to the temperature of the
thermal bath. For a recent discussion based on solutions of the
Boltzmann equation relevant during the warm inflation dynamics, see,
e.g., Ref.~\cite{Bastero-Gil:2017yzb}. In the following we assume thermal equilibrium with $T_{\delta \phi} = T$. Typically, $G(Q)$ reduces to $1$ for $Q=0$ and in most of the known models it is well approximated by a fraction of polynomials in $Q$ with numerically fitted coefficients~\cite{Graham:2009bf, BasteroGil:2011xd}. Notice that for $Q=0$, which also implies $T = T_{\delta \phi} = 0$, we recover the usual cold inflation spectrum given in Eq.~\eqref{eq:vacuum_contribution} as expected for consistency. 

As a matter of fact the presence of radiation is thus inducing a series of modifications in the typical CMB observables, namely $n_s$ and $r$. In particular, we expect two competing effects:
\begin{enumerate}
  \item The decay of the inflaton into radiation is effectively playing the role of an additional friction term for the inflaton beyond the usual Hubble friction. As a consequence we expect, similarly to~\cite{Domcke:2016bkh}, a shift in the point of the potential probed by CMB observations. In particular this effect is expected to produce a decrease of $n_s$ and an increase in $r$.
  \item The radiation will play the role of a source term for scalar field fluctuations which induces an amplification in the scalar power spectrum. Indeed, this can be noticed by Eq.~\eqref{eq:modified_scalar}. However, in general we do not expect a similar coupling between radiation and tensor fluctuations. As a consequence this effect induces an increase in $n_s$ and a decrease in $r$.
\end{enumerate}
If thermal effects are already important (or at least not completely negligible) at CMB scales, the first of these two effects happens to be subdominant with respect to the second one, meaning that typically $n_s$ is increasing and $r$ is decreasing with respect to the cold case.

For completeness, we should mention that an analysis of non-Gaussianities has been performed for  warm inflation for the weak and strong dissipation regimes, see, e.g., Refs.~\cite{Moss:2007cv,Bastero-Gil:2014raa}. In both cases the predictions are generally in good agreement with the Planck constraints in Ref.~\cite{Ade:2015ava}.

\section{$\beta$-function formalism and warm inflation}
\label{sec:warm_beta}

Let us start this section with a brief review of the main ideas behind
the formulation of the $\beta$-function formalism for inflation
introduced in Ref.~\cite{Binetruy:2014zya}. In the simplest
realization of inflation, i.e. a single field, with a
standard kinetic term and no interactions with other particles, the
evolution of the Universe during inflation is completely specified by the system of equations
\begin{eqnarray}
	\label{eq:einstein_cold}
	3H^2 &=& \rho \; , \\
  \label{eq:Raychauduri}
  -2 \dot{H} &=& p +\rho \; , \\ 
	\label{eq:klein_gordon_cold}
	\qquad \ddot{\phi} + 3H \phi +V_{,\phi} &=& 0 \;  , 
\end{eqnarray}
i.e., by the Einstein equations and the equation of motion for the
scalar inflaton $\phi$. In the spirit of the HJ formalism, the
solution of this system is assumed to exist and, in particular, the time
evolution of the inflaton field $\phi$  is
assumed to be piecewise monotonic. Under this assumption it is
possible to invert $\phi(t)$ to get $t(\phi)$ and the field can be
directly used as the clock describing the evolution of the system.

By introducing the so-called superpotential\footnote{This choice for
  the name is justified by the formal analogy with the
  parameterization of the scalar potential in SUSY (for a review, see
  for example, Ref.~\cite{Binetruy:2006ad}).} $W(\phi)\equiv-2H(\phi)$
and using Raychauduri's equation~\eqref{eq:Raychauduri} we
easily obtain that
\begin{align}
	\dot{\phi}=W_{,\phi}\;. \label{eq:phidotphi_cold}
\end{align}
This equation clearly shows that by exploiting the HJ formalism it is
possible to express $\dot{\phi}$ as a function of $\phi$ only. The $\beta$-function is then defined as
\begin{equation}
	 \label{eq:beta_function_cold_case}
	 \beta(\phi) \equiv \frac{\textrm{d} \phi}{\textrm{d} \ln a} =
         \frac{\dot{\phi}}{H} = - 2 \frac{W_{,\phi}}{W} \; .
\end{equation}
Notice that, in analogy with the RGE, the scalar field is playing the role
of the coupling constant and the scale factor is playing the role of
the renormalization scale. Using the definition of
Eq~\eqref{eq:beta_function_cold_case}, the equation of state becomes
\begin{equation}
	 \label{eq:eq_of_state_cold}
	 \frac{p+\rho}{\rho} = \frac{4}{3} \, \frac{W_{,\phi}^2 }{
           W^2} = \frac{\beta^2 (\phi)}{3} \; .
\end{equation}
This equation implies that an exact dS geometry corresponds to a fixed point with $\beta(\phi) = 0$ and thus a phase of nearly exponential expansion of the Universe is realized by departing from a region where
$\beta \ll 1$. As a consequence, in the framework of the
$\beta$-function formalism, a class of models of inflation is defined by specifying the asymptotic parameterization of
$\beta(\phi)$ in the neighborhood of a point where $\beta(\phi) \ll
1$. Moreover while the leading contribution around the fixed point
  sets the main properties of the universality class, higher order
  contributions to $\beta(\phi)$, which are negligible in the
  neighborhood of fixed point, break the universality at different
  scales.

For warm inflation a model is not only specified by the
inflationary potential, but also by the dissipation coefficient ratio
$Q$, which in general is a function of both $\phi$ and $T$. Once
these two functions are specified, the evolution is completely determined by the set of
equations~\eqref{eq:kleingordon}-\eqref{eq:raychauduri}. By
solving these equations, we can express all the relevant quantities, i.e., $H(t),\ \phi(t),\ Q(t)$, and $T(t)$, as functions of time. Once again the
problem can be studied in the framework of the HJ formalism and,
assuming the evolution of $\phi(t)$ to be piecewise monotonic, it is
possible to compute, at least locally, $t(\phi)$ and express all
the relevant quantities as functions of the field only.

In analogy with the treatment carried out in the cold case, we introduce a superpotential $W(\phi)\equiv-2H(\phi)$. Assuming the radiation energy density to be quasi-stable\footnote{More on this approximation is said below Eq.~\eqref{eq:quasi_stable_beta} where we re-express this condition in terms of the typical quantities of the formalism.}, meaning
$\dot{\rho}_r\ll4H\rho_r$, and using the Raychauduri's equation~\eqref{eq:raychauduri}
we  thus obtain
\begin{align}
	\dot{\phi}=\frac{W_{,\phi}}{1+Q}\;. \label{eq:phidotphi}
\end{align}
By using this equation and Eq.~\eqref{eq:raychauduri} we get from Eq.~\eqref{eq:p_phi_rho_phi_rho_T}
\begin{align}
	T^4=\frac{45}{2\pi^2
          g_*}\frac{Q}{(1+Q)^2}W_{,\phi}^2\;.\label{eq:Tempofphi}
\end{align}
To find the temperature as a function of $\phi$ only, Eq.~\eqref{eq:Tempofphi} needs to be solved for $T$. Note that
since in general $Q$ depends both on $T$ and $\phi$, the solution of
this equation might exist only numerically. Then, once $T(\phi)$ is known,
the dissipation coefficient ratio is expressed as a function of $\phi$
only\footnote{In principle, it could also be possible to start by
  directly fixing a parameterization for $\mathcal{Q}(\phi)$. More on
  this will be commented in
  Sec.~\ref{sec:conclusions_and_future}.} as $Q(T(\phi),\phi)\equiv
\mathcal{Q}(\phi)$. Note that a different notation, $\mathcal{Q}$, is
used here to stress the difference in the functional
dependence on $\phi$.  

We proceed our discussion by introducing the cosmological
$\beta$-function as defined in Eq.~(\ref{eq:beta_function_cold_case}),
$\beta(\phi) \equiv \textrm{d} \phi/\textrm{d} \ln a=\dot{\phi}/H$.
Note that the analogy with a RGE still holds. The equation of state reads
\begin{align}
	-\frac{2\dot{H}}{3H^2} =\frac{(1 + Q)
          \dot{\phi}^2}{3H^2}=(1+Q)\frac{\beta^2(\phi)}{3} \;. \label{eq:eqofstate1}
\end{align}
Interestingly, Eq.~\eqref{eq:eqofstate1} shows that an exact dS
geometry is again realized in correspondence to the zeros of
$\beta(\phi)$ and the phase of accelerated expansion of
the Universe stops when $(1+Q)\beta^2(\phi)$ is of order one. This
is a crucial difference with respect to the cold case in Eq.~\eqref{eq:eq_of_state_cold}. As $Q$ is always positive, the fixed point
is only attained by a vanishing $\beta$-function, but in general, unless we have $\beta$ exactly equal to zero,
$\beta^2(\phi) \ll 1$ is not sufficient to ensure that the Universe is inflating. In
particular the Universe may stop to inflate because $(1+Q) \gg
\beta^{-2}(\phi)$, while $\beta \ll 1$. Another original and strictly warm realization of inflation is the case in which, departing from the dS fixed
point, $\beta(\phi)$ reaches a constant value smaller than one. In
such a scenario, the last part of the inflationary phase is thus
driven and, in particular, is concluded by the evolution of $Q$. As inflation can only be realized for
$\beta(\phi) \ll 1$, its parametrization can still be used to fix the
flow in the neighborhood of the fixed point. Once again it is thus
possible to use $\beta(\phi)$ to define a set of universality classes
as in the cold inflation case.

To make the generalization from cold inflation more evident, let us
define
\begin{align}
 	 \beta_{CI}(\phi)& \equiv -2\frac{W_{,\phi}}{W}=(1+Q)\beta(\phi)\;,
 	 \label{eq:beta_ci}
 \end{align}
 which has the exact same dependence on $W$ as the beta function of
 the cold inflation,  Eq.~(\ref{eq:beta_function_cold_case}). Note
 that with this definition, the superpotential $W$ can be readily
 expressed as
 \begin{equation}
 	W(\phi)=W_{\textit{f}}\exp\left[-\frac{1}{2}\int_{\phi_{\textit{f}}}^\phi
         \textrm{d} \phi'\beta_{CI}(\phi')\right]\;, \label{eq:superpotfrombeta}
 \end{equation}
 where the subscript $\textit{f}$ is used to denote quantities evaluated at the end of inflation.
Moreover, using the definition given in Eq.~\eqref{eq:beta_ci}, it is
easy to prove that the equation of state can be expressed as
\begin{align}
	-\frac{2\dot{H}}{3H^2} &= \frac{\beta_{CI}^2(\phi)}{3(1+Q)}\;.
\end{align}
Again, the fixed point is reached when $\beta_{CI}$ goes to zero and we
 see that inflation ends when $\beta_{CI}^2\sim1+Q$. Notice that according to Eq.~\eqref{eq:superpotfrombeta}
$\beta_{CI}$ is directly associated with the superpotential and thus
with the inflationary potential. This equation makes clear that for
$Q$ sufficiently large, the Universe is inflating for $\beta_{CI} \gg
1$. In this sense, the dissipation coefficient can be interpreted as a
friction term that slows down the evolution of the inflaton field
and this is potentially allowing for inflation in regions of the
potentials that are steeper than the ones usually considered in the
cold case. As already mentioned in the previous section, this could provide a mechanism to ease the $\eta$-problem. In order to generalize the universality classes defined for
cold inflation in Ref.~\cite{Binetruy:2014zya}, in this work we will
simply use the $\beta$-functions associated with these classes as
choices for $\beta_{CI}$. We then observe how the different
dissipation coefficient  ratios will affect the predictions of any classes, this analysis is carried out in Sec.~\ref{sec:methods_and_results} and \ref{sec:discussion_of_results}.

At this point we can translate the  quasi-stable assumption of the  radiation energy density in the language of the $\beta$-function formalism
\begin{equation}
\label{eq:quasi_stable_beta}
 \left| \frac{\beta}{4 } \frac{\textrm{d} \ln \rho_r}{\textrm{d} \phi} \right| = \left| \frac{\beta_{CI}}{4(1 + Q) }  \left[ \frac{Q_{,\phi}}{Q} \left( \frac{1 - Q }{1 + Q } \right)  + 2 \frac{\beta_{CI,\phi}}{\beta_{CI}} - \beta_{CI} \right]\right| \ll 1 \;.
\end{equation}
The validity of this condition  has to be checked for each choice of $\beta_{CI}$ and $Q$. However it is possible to show that for all the cases discussed in this paper, this assumption is satisfied.

{}The expression of the number of e-foldings $N$ in this formalism reads
\begin{equation}
	N(\phi)\equiv-\int_{a_{\textit{f} } }^a \textrm{d} \ln a= -\int_{\phi_{\textit{f}}}^\phi 	\frac{ \textrm{d} \phi'
 }{\beta (\phi')} = -\int_{\phi_{\textit{f}}}^\phi  \textrm{d} \phi'
	\frac{1+Q(\phi')}{\beta_{CI}(\phi')}\;,\label{eq:numberofefoldings}
\end{equation}
where $\phi_{\textit{f}}$ is the field value at the end of inflation, fixed by
$\beta_{CI}^2(\phi_{\textit{f}})= 1+Q(\phi_{\textit{f}}) $, and the expression of the
inflationary potential which is derived by using
Eq.~\eqref{eq:friedmann},
 \begin{align}
 	V(\phi)&=\frac{3}{4}W^2(\phi)\left[1-\frac{1}{6}\frac{(1 + 3Q/2 )}{(Q+1)^2}
 	\beta_{CI}^2(\phi)\right]\;. \label{eq:potentialfrombeta}
 \end{align}
As for the physically relevant cases, we expect both $\beta_{CI}$ and
$Q$ to be negligible while the Universe is deep into the inflationary
phase, i.e., for large values of $N$, the parameterization of the
inflationary potential is typically mainly determined by the
superpotential $W(\phi)$. {}It is worth mentioning that the
formalism is not only valid at the background level, but rather it can
also be used to describe cosmological perturbations.

\begin{figure}[htb!]
 {\includegraphics[width=0.5\columnwidth]{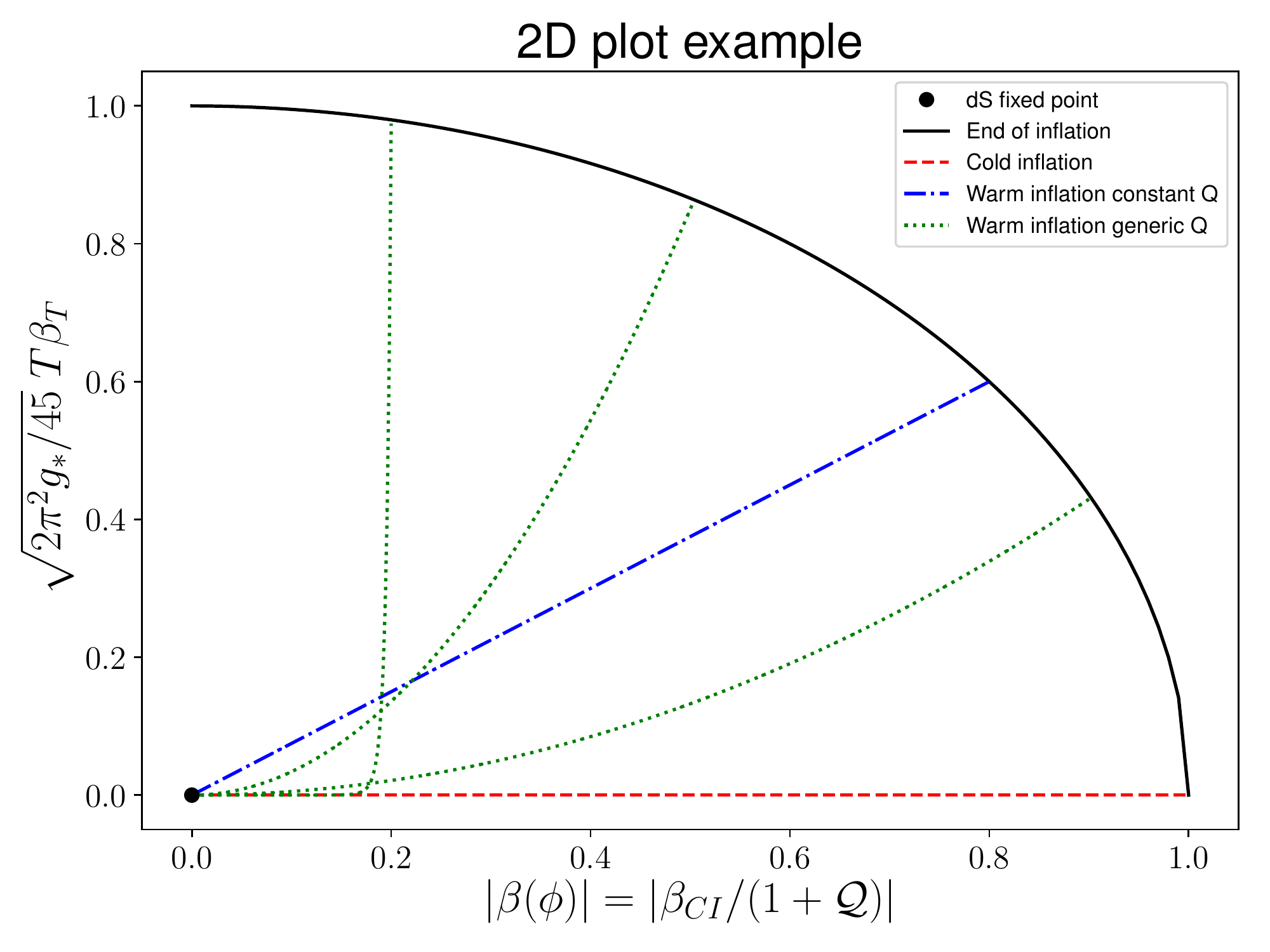}}
\caption{Some possible inflationary trajectories corresponding
   to the flow from the dS fixed point to the solid black
  line and  showing the departure from the usual cold inflation case (red dashed line). The curves shown in this plot are illustrative examples which
  are not corresponding to any concrete model. \label{fig:2D_example}}
\end{figure}

To grasp a better understanding of the competing influences of
$\beta_{CI}$ and $Q$ during the phase of inflation, it is worth
defining the  complementary function $\beta_{T}$ as
\begin{align}
	\label{eq:beta_temperature}
	\beta_{T}(\phi) \equiv \frac{T}{H} = -2 \frac{T}{W}\;.
\end{align}
Using Eq.~\eqref{eq:p_phi_rho_phi_rho_T} we express
\begin{equation}
	\label{eq:rho_r_over_H}
	\frac{\rho_r}{H^2} =  \frac{\pi^2 g_*}{30} \frac{T^4}{H^2} =
        \frac{\pi^2 g_*}{30} \left[ T \beta_T(\phi)\right]^2 \; , 
\end{equation}
which makes manifest the interpretation of $\beta_{T}$. This function
captures the amount of radiation produced during warm inflation. In
particular, by considering the full equation of state,
\begin{eqnarray}
	-\frac{2\dot{H}}{3H^2}
        &=&\frac{\dot{\phi}^2+\frac{4}{3}\rho_r}{3H^2}  \nonumber
        \\ &=&\frac{\beta^2 (\phi)}{3} + \frac{2\pi^2 g_*}{45}
        \frac{\left[ T \beta_T(\phi)\right]^2}{3}  \nonumber
        \\ &=&\frac{1}{3}\left[ \frac{\beta_{CI}(\phi)}{Q+1}\right]^2
        + \frac{2\pi^2 g_*}{45} \frac{\left[ T
            \beta_T(\phi)\right]^2}{3}\; , 
	\label{eq:eqofstate2}
\end{eqnarray}
it is clear that (as $T \neq 0$) $\beta_T$ parameterizes the flow from
the dS fixed point induced by radiation. Interestingly, using the
definitions of $\beta_{CI}$ and $\beta_{T}$ we can represent the phase
of inflation in a two dimensional plot depicting the departure from the
usual cold inflation case. In Fig.~\ref{fig:2D_example} the phase of inflation is
represented as a trajectory starting from, or close to, the dS fixed
point at the origin and reaching the circle of unitary radius, where
$(1 + Q_\textit{f})\beta_\textit{f}^2 = 1$, which corresponds to the end of inflation.  From the equation of state~\eqref{eq:eqofstate2} we note that the axes in Fig.~\ref{fig:2D_example}  are proportional the square roots of the fractional kinetic and thermal energy densities. The flow along the different inflationary trajectories can be directly
parametrized by the value of the inflaton field $\phi$, or
equivalently by the number of e-foldings $N$ defined in
Eq.~\eqref{eq:numberofefoldings}. A motion in the horizontal
direction is due to $\beta_{CI}$, whereas a vertical motion is an
effect of production of radiation. Since we have that
\begin{align}
	T\beta_{T}(\phi)=\sqrt{\frac{45}{2\pi^2g_*}\frac{Q}{(Q+1)^2}}|\beta_{CI}(\phi)|
        \;,\label{eq:Tbetaofphi}
\end{align}
we observe that the shape of the trajectory is mostly defined by the
dissipation coefficient ratio $Q$. Note that in these kind of plot, any model of cold inflation is
represented as an horizontal line with $T \beta_T = 0$. Conversely, warm inflation models are expected to be
represented as curves departing from this line. As all the
inflationary trajectories are expected to end on the solid black line,
large values of $\beta_\textit{f}$, i.e., for values closer to
$T_\textit{f} \beta_{T \, \textit{f}} = 0$, imply a small radiation
contribution to the equation of state at the end of
inflation. Conversely, small values for $\beta(\phi_\textit{f})$,
imply a non-negligible radiation contribution to the equation of state
at the end of inflation. 

Apart from the de Sitter fixed point at the origin, there are two other special points on Fig.~\ref{fig:2D_example}. 
The first is (1,0), where cold inflation usually ends. When a trajectory crosses this point, the Universe stops 
inflating and it must then enter into the (p)reheating phase. The second point, which is not appearing on our plots, 
should be (0,2). This corresponds to the Universe being in the radiation-dominated era, i.e. $\rho_r/(3H^2)\simeq1$. 
Notice that all trajectories describing viable cosmological models, which consistently include the 
evolution of the Universe after inflation, must cross this point. However, since in (0,2) we have 
$\dot{\rho}_r = 4 H \rho_r$, for sure the assumption of quasi-stable radiation must be violated, 
implying that our treatment cannot be extended all the way up to this point.

A model which touches the solid black line
for large (order 1) values of $\sqrt{2\pi^2g_* /45} \, T \beta_T$, for example the vertical dotted curve in {}Fig.~\ref{fig:2D_example},
implies that the RG flow in the last part of inflation is mainly
induced by the radiation. This does not imply that the Universe is
dominated by the radiation, but rather that it is rapidly approaching
the moment where the transition from inflation to a radiation
dominated Universe takes place. Since in these models the radiation
energy density at the end of inflation is already sizable and the
inflaton kinetic energy is small, an explosive (p)reheating may not
be required and the transition from inflation to radiation may be smooth.  This has to be checked model by model. Since in warm inflation it is possible to unify the treatment of inflation and (p)reheating, a self-consistant 
computation of $N_{CMB}$, the value of $N$ at which CMB observables leave the horizon, could in principle be 
carried out. However, in order to perform this analysis, we need to study the trajectory until the point (0,2) 
is reached, which requires to violate the assumption of quasi-stable radiation. As this goes beyond the scope of 
this work, where required to adopt $N_{CMB}=60$ as a representative value.

For completeness the remaining relevant quantities in the description of inflation expressed in terms of the $\beta$-function formalism can be found 
Appendix~\ref{sec:appendix_formula}, in particular, the scalar-spectral index and the tensor-to-scalar ratio.

\section{Applying the formalism to explicit examples}
\label{sec:methods_and_results}
In this section we provide a general procedure for
computing the predictions in the $\beta$-function formulation of warm
inflation. In particular, we start by explaining our
numerical methods for examining models,
and  then focus on some special cases
that admit an analytical treatment. As already explained in
Sec.~\ref{sec:warm_beta}, the model is completely
specified by fixing a $\beta$-function, either $\beta(\phi)$ or
$\beta_{CI}(\phi)$, and by a dissipation coefficient ratio
$Q(T,\phi)$. In order to generalize the classes of universality for
cold inflation~\cite{Binetruy:2014zya}, we choose to start by
fixing a parameterization for $\beta_{CI}(\phi)$. The dissipation
coefficient  $\Upsilon(T,\phi)$ is derived explicitly by QFT methods, see, e.g.,
Refs.~\cite{BasteroGil:2010pb,BasteroGil:2012cm}. In this work, we
focus on a rather general parameterization for the dissipation ratio
$Q=\Upsilon/(3H)$ that is motivated by the previous warm inflation
models developed in the literature 
\cite{Berera:1998px,Bastero-Gil:2016qru,Bartrum:2013fia,BasteroGil:2010pb,BasteroGil:2012cm,Zhang:2009ge},
\begin{align}
	\label{eq:Q_general_parameterization}
	Q&=\frac{CT^m}{H\phi^n}=-\frac{2CT^m}{W\phi^n}\;,
\end{align}
where $C$ is a constant. This example will also facilitate the illustration of the methodology. When a
complete specification of the model is required, i.e. an explicit
choice for $\beta_{CI}(\phi)$, for simplicity we will restrict our analysis to the chaotic
class 
\begin{align} 
	\beta_{CI}(\phi) = -\frac{\alpha}{\phi}\;, \label{eq:chaoticbetaci}
\end{align}
where $\alpha$ is a positive constant. The generalisation to other
classes of models can be carried out analogously.

In general it is unlikely to have a complete analytical description
of the model and therefore numerical methods are required. The procedure we have used to derive numerical solutions is the
following: having $\beta_{CI}(\phi)$, $Q(T,\phi)$ and an initial guess
value for the constant $W_{\textit{f}}$, which fixes the normalization of the
inflationary potential, as  inputs, the value of the scalar field at
the end of inflation $\phi_{\textit{f}}$ and the corresponding temperature $T_{\textit{f}}$
are computed using 
\begin{equation}
\begin{aligned}
	\beta^2_{CI,\textit{f}} &= 1 + Q_\textit{f}  \;, \\
        T_\textit{f}^4 & = \frac{45}{8 \pi^2 g_*}
        \frac{Q_\textit{f}}{1+Q_\textit{f} } W_\textit{f}^2
        \;. \label{eq:phiendandTend}
\end{aligned}
\end{equation}
where we recognize Eq.~\eqref{eq:Tempofphi} in second of these equations. They can be recasted as
\begin{align}
	T_\textit{f} &=  \left( \frac{45 C}{4 \pi^2 g_*} \frac{-
            W_\textit{f}}{\beta^2_{CI,\textit{f}} \phi_\textit{f}^n }
          \right)^{\frac{1}{4-m}} \; ,\\ 
          \beta^2_{CI,\textit{f}} &= 1
        + \left( \frac{45}{8 \pi^2 g_* \beta^2_{CI,\textit{f}} }
          \right)^{\frac{m}{4-m}}  \left( \frac{2C}{\phi_\textit{f}^n}
        \right)^{\frac{4}{4-m}}\left( -  W_\textit{f}
        \right)^{\frac{2m-4}{4-m}}\;.
\end{align}
The solution of the above system of equations is obtained by first
solving for $\phi_\textit{f}$ and then computing $T_{\textit{f}}$. 
The inflaton field then serves as a clock for the evolution of the system. 
We evolve the field from $\phi_\textit{f} $ to $ \phi_\textit{f} \pm \Delta \phi$
with $\Delta \phi \ll \phi_\textit{f}$ being an infinitesimal
step. The sign of the increment is fixed by the position of the fixed
point, i.e., whether the value of the field increases or
decreases during inflation. At this point the relevant quantities $T$, $Q$ and
$N$ are evaluated at $\phi_\textit{f} \pm \Delta \phi$ using Eq.~\eqref{eq:Tempofphi}, the definition
of $Q$ and Eq.~\eqref{eq:numberofefoldings}, respectively. The procedure is then repeated until the value $\phi_{CMB}$ is reached. The latter is defined as the value of $\phi$ which gives $N(\phi_{CMB})=N_{CMB}$ where in this work we assume $N_{CMB} = 60$ as the value of $N$ at which CMB observables leave the horizon. As a consequence, the evolution is solved for all the scales between the end of inflation and CMB scales. Finally, by comparing the amplitude of the scalar power spectrum
with the COBE normalization~\cite{Planck2015,Ade:2015xua}, it is
possible to adjust the constant $W_{\textit{f}}$ in order to satisfy this
constrain. The predictions for the scalar spectral index and
tensor-to-scalar ratio are then computed from
Eqs.~\eqref{eq:nswarminfl} and~\eqref{eq:rwarminfl} for the values of
$\phi$ corresponding to $N_{CMB}$. These quantities can
finally be compared with the observational
constraints~\cite{Planck2015,Ade:2015xua}.

\subsection{Analytical Methods}
\label{sec:analytical_methods}
In this subsection we focus on some cases where a complete (or
partial) analytical treatment can be performed. In order to carry out
this treatment we have to
\begin{enumerate}
	\item Compute $\phi_{\textit{f}}$ and $T_{\textit{f}}$, the values of the inflaton field and of the temperature at the end of inflation using Eqs.~\eqref{eq:phiendandTend};
	\item Derive the superpotential and its derivative using
          $\beta_{CI}$ and Eq.~\eqref{eq:superpotfrombeta};
	\item \label{step:critic_step} Compute $T(\phi)$ by solving
          Eq.~\eqref{eq:Tempofphi} with the dissipation coefficient ratio
          $Q(T,\phi)$ written explicitly in terms of $T$ and
          $\phi$. Having $T$ as a function of the field, we can also
          write $Q(T(\phi),\phi)$ as a function of $\phi$ only;
	\item {}Finally, we express $\beta_T$ as a function of $\phi$
          using Eq.~\eqref{eq:Tbetaofphi}.
\end{enumerate}
Note that in general, the third step cannot be
carried out analytically for non-trivial forms of the dissipation
coefficient. Typically, it is also useful to derive all the relevant
quantities as functions of the number of e-folds. {}For this purpose,
we thus compute the number of e-folding $N(\phi)$ from
Eq.~\eqref{eq:numberofefoldings} and invert it to find $\phi(N)$. Note
also that once again we fix the constant $W_{\textit{f}}$ in order to be
consistent with the COBE normalization $\Delta_s^2(N_{CMB}=60)=2.2\times
10^{-9}$. In particular this is done by solving
Eq.~\eqref{eq:scalarspectrumbeta} for $W(N)$ at $N=N_{CMB}$. Let us now
illustrate the method with some examples where a partial (or complete) analytical
treatment exists.

\subsubsection{Constant Q - Full analytical treatment}
\label{sec:constant_analytic}
Let us restrict to the simplest case possible, that of a constant
dissipation coefficient ratio $Q(T,\phi)=Q$. We first consider a generic $\beta_{CI}$  and then restrict to the specific example of the chaotic
class specified by Eq.~\eqref{eq:chaoticbetaci}. For a constant $Q$,
Eq.~\eqref{eq:Tempofphi} admits the solution
\begin{equation}
	\label{eq:T_constant_solution}
	T(\phi)=\left[ \frac{45}{8\pi^2g_{*}}\frac{Q}{(1+Q)^2} W^2(\phi)
          \beta_{CI}^{ \ 2}(\phi) \right]^{1/4}\;,
\end{equation}
where $W(\phi)$ is directly set by Eq.~\eqref{eq:superpotfrombeta}. To
check the consistency of the model, we can compute $\rho_r$, by
substituting Eq.~\eqref{eq:T_constant_solution} into
Eq.~\eqref{eq:rho_r_over_H},
\begin{equation}
	\label{eq:rhor_constantq}
	\rho_r = \frac{3}{16 }\frac{Q}{(1+Q)} W^2(\phi) \beta_{CI}^{
          \ 2}(\phi) \; .
\end{equation}
Interestingly, this can be compared with the result
\begin{equation}
	\rho_\phi = \frac{3}{4}W^2 - \rho_r = \frac{3}{4}W^2 \left[ 1
          - \frac{Q}{(1+Q)} \frac{\beta_{CI}^{ \ 2}(\phi)}{4}\right]
        \; ,
\end{equation}
to conclude that, independently on the value of $Q$, when we approach
the dS fixed point, $\beta_{CI}(\phi) \ll 1$, we always consistently
get $\rho_r \ll \rho_\phi$. 

To further proceed we need to precise a parameterization for $\beta_{CI}$ and therefore we restrict ourselves to the case of
the chaotic class of Eq.~\eqref{eq:chaoticbetaci}. In this case, the
superpotential and the temperature, respectively, read
\begin{align}
	W &= W_{\textit{f}}
        \left(\frac{\phi}{\phi_\textit{f}}\right)^{\frac{\alpha}{2}}\;,
        \\ T(\phi)&= \left[\frac{45}{8\pi^2g^*}\frac{Q}{(1+Q)^2} \frac{\alpha^2 W_\textit{f}^2}{\phi_\textit{f}^\alpha }  \phi ^{\alpha-2}\right]^{1/4} \;.
\end{align}
{}For completeness, we also derive, using
Eq.~\eqref{eq:potentialfrombeta}, the potential
\begin{equation}
	V = \frac{3}{4} W_{\textit{f}}^2
        \left(\frac{\phi}{\phi_\textit{f}}\right)^{\alpha} \left[ 1 -
          \frac{\alpha^2}{12 \phi^2} \frac{2 + 3 Q}{(1 + Q)^2}\right]
        \simeq \frac{3}{4} W_{\textit{f}}^2
        \left(\frac{\phi}{\phi_\textit{f}}\right)^{\alpha} \; ,
\end{equation}
where the approximation in the last step relies on $\phi \gg \alpha$, which is valid deep in the
inflationary phase. The value of the field at the end of inflation is
\begin{align}
	\phi_{\textit{f}}&=\sqrt{\frac{\alpha^2}{1+Q}}\;,
\end{align} 
and the number of e-foldings $N$ as a function of $\phi$ reads
\begin{equation}
	N = \frac{1+Q}{2\alpha} \left( {\phi^2}- \frac{\alpha^2}{1+Q}
        \right) \; , 
\end{equation}
which implies
\begin{equation}
   \phi = \sqrt{\frac{2\alpha N+\alpha^2}{1+Q}}  \; .
\end{equation}
At this point we can also compute $\beta_{CI}(N)$, $T(N)$ and
$\beta_{T}(N)$, whose expressions are given, respectively, by
\begin{align}
  	\beta_{CI}(N) & =-\sqrt{\frac{(1+Q)\alpha}{2N+\alpha}} \;,
 	 \\ T(N) & =\left[ \frac{45}{8\pi^2g^*}\frac{Q}{ 1+Q } \alpha^{2-\alpha} W_\textit{f}^2   \left( 2 \alpha N + \alpha^2  \right)^{\alpha/2 - 1} \right]^{1/4} \;,\\
 	 \beta_T(N)&= \left[\frac{90}{\pi^2g_*}\frac{Q}{1+Q}
	  \frac{\alpha^{2+\alpha}}{W_{\textit{f}}^2}  \left( 2 \alpha N + \alpha^2  \right)^{-\alpha/2 - 1}  \right]^{1/4}\;.
\end{align}
Notice that for $\alpha=2$ the temperature is constant during
inflation. {}Finally, we fix $W_{\textit{f}}$ using the COBE normalization
and we compute the spectral tilt $n_s$ and the tensor-to-scalar
ratio $r$ using Eqs.~\eqref{eq:nswarminfl}
and~\eqref{eq:rwarminfl}. As $Q$ is positive, we expect a slightly
increased value of $n_s$ and a slightly reduced value of $r$ with
respect to the cold inflation case.

\subsubsection{Weak and strong dissipative limits}
\label{sec:weak_strong_dissipation}
{}For the general choice of $Q(T,\phi)$,
Eq.~(\ref{eq:Q_general_parameterization}), an analytical description
does not exist in all regimes. However, similarly to the treatment of
Ref.~\cite{Sayar:2017pam}, an analytical description of these models
can be achieved both in the strong $Q\gg1$ and in the the weak $Q\ll
1$ dissipative limits. In particular, it is possible
to derive analytical expressions for $\mathcal{Q}$ and $T$ as
function of $\phi$ only, which we do next.

\paragraph{Weak dissipative regime} Let us consider the
parameterization of $Q(T,\phi)$ given in
Eq.~\eqref{eq:Q_general_parameterization}. In the limit $Q\ll1$
we can immediately use Eq.~\eqref{eq:Tempofphi} to compute the
temperature to obtain
\begin{equation}
	T(\phi)=\left[\frac{45C}{4\pi^2 g_*
        }\frac{\beta_{CI}^2(-W)}{\phi^n}\right]^{\frac{1}{4-m}}\;,
\end{equation}
and then, by substituting this expression into
Eq.~\eqref{eq:Q_general_parameterization}, we find
\begin{align}
	\mathcal{Q}(\phi)  &=2C\left(\frac{45C}{4\pi^2 g_*
         }\right)^{\frac{m}{4-m}}
         \left(-W\right)^{\frac{2(m-2)}{4-m}}\phi^{\frac{-4n}{4-m}}
         \beta_{CI}^{\frac{2m}{4-m}}\;.
\end{align}
To completely specify the model, we need to substitute an explicit parameterization for
$\beta_{CI}$. {}For the example of chaotic class Eq.~\eqref{eq:chaoticbetaci}, we find that
\begin{align}
 	 T(\phi)&\simeq\left[\frac{45C\alpha^2
   	 (-W_{\textit{f}}\phi_{\textit{f}}^{-\frac{\alpha}{2}})}{4\pi^2g_*}\right]^{\frac{1}{4-m}}
 	 \phi^{\frac{\alpha-4-2n}{2(4-m)}}\;,\\
	  \mathcal{Q}(\phi)
 	 &\simeq2C\left(\frac{45C\alpha^2}{4\pi^2
          g_* }\right)^{\frac{m}{4-m}}
 	 \left({W}^2_f{\phi_{\textit{f}}^{-\alpha}}\right)^{\frac{(m-2)}{4-m}}
 	 \phi^{\frac{\alpha( m-2)-2m-4n}{4-m}}\;.
\end{align}
It is worth to note that this regime can only be attained dynamically for a certain
set of values for $\alpha$, $C$, $n$ and $m$. Recall that $W_{\textit{f}}$ is fixed by the
COBE normalization and, thus, it should not be considered a free
parameter. In particular, as the chaotic class describes large field
models, meaning that inflation takes place for large values of
$\phi$, this can only be attained if ${\frac{\alpha(
    m-2)-2m-4n}{4-m}} < 0$. It is interesting that since in the chaotic class both the superpotential and the $\beta$-function have the form of a power law, the temperature and $\mathcal{Q}(\phi)$ must have a power law dependence as well. This behavior can actually change for different
classes\footnote{\label{foot:different_behaviors} For both the monomial
  and inverse classes (see Eq.~\eqref{eq:monomial_class} and
  Eq.~\eqref{eq:inverse_class}) $\beta_{CI}(\phi)$ is still a power
  law, but $W(\phi)$ are respectively given by
  Eq.~\eqref{eq:monomial_superpot} and
  Eq.~\eqref{eq:inverse_superpot}. As a consequence an approximate
  power law behavior can only be attained in regions where $W$ is
  nearly constant, i.e., where $\phi$ is very close to the fixed point
  (meaning deep in the inflationary phase). Conversely, for the
  exponential class (see Eq.~\eqref{eq:exponential_class}) the
  $\beta$-function is not a power law and, thus, the power law
  behavior is never approached.}. The dependence of $T$ and $\mathcal{Q}$ on $\phi$ for some particular
choices of $m$ and $n$ are written in Tab.~\ref{tab:PLBchaoticweak}.
\begin{center}
\captionof{table}{Power-law behaviors of $\mathcal{Q}(\phi)$ and $T(\phi)$ for the chaotic class in the weak dissipative limit.}
\begin{tabular}{|l|c|c|}
	\hline
	\textbf{Dissipation Coefficient Ratio} & $\mathcal{Q}(\phi)\sim \phi ^ \#$ & $T(\phi)\sim \phi ^\#$\\
	\hline
	\text{Cubic ($m=3$, $n=2$) } & $ {\alpha-14}$ & ${(\alpha-8)/2}$ \\
	\text{Linear ($m=1$, $n=0$) } & $ {-(\alpha+2)/3}$ & ${(\alpha-4)/6}$ \\
  	\text{Inverse ($m=-1$, $n=0$) } & $ {(-3\alpha+2)/5}$ & ${(\alpha-4)/10}$	 \\
	\hline
\end{tabular}
\label{tab:PLBchaoticweak}
\end{center}
A graphic representation of these behaviors for particular sets of $m$
and $n$ can be seen in {}Fig.~\ref{fig:asymptotic} shown in Sec.~\ref{sec:discussion_of_results}. Assuming that the
model stays in the weak dissipative regime (this assumption has to be
checked model by model) for the whole period of inflation, we can
proceed further with the computation of the number of e-foldings,
\begin{align}
	\label{eq:N_of_phi_weak}
	N(\phi)&=\frac{1}{2\alpha}\left(\phi^2-\alpha^2\right)\;.
\end{align}
Using Eq.~\eqref{eq:phiendandTend}, it is now possible to compute the
value of the inflaton field at the end of inflation as given by
$\phi_{\textit{f}}=\alpha$. At this point, in order to check the consistency of
the approximation, we should verify that $\mathcal{Q}(\phi_{\textit{f}})\ll1$ or
\begin{align}
	2C\left(\frac{45C}{4\pi^2 g_* }\right)^{\frac{m}{4-m}}
        \left(- {W}_f \right)^{\frac{2(m-2)}{4-m}}\alpha^{\frac{-4n}{(4-m)}}\ll1\;.
\end{align}
Finally, by inverting Eq.~\eqref{eq:N_of_phi_weak}, we obtain
\begin{align}
	\phi(N)&=\sqrt{2\alpha N+\alpha^2}\;.
\end{align}
{}Having derived $Q(\phi)$, $T(\phi)$ and $\phi(N)$, we can
immediately compute the predictions for $n_s$ and $r$ using
Eq.~\eqref{eq:nswarminfl} and Eq.~\eqref{eq:rwarminfl}.
 
\paragraph{Strong dissipative limit} Let us follow a procedure for the 
strong dissipative regime analogous to the one carried out above for the 
weak dissipative limit. As a first step, we compute the temperature and the dissipation 
coefficient as functions of $\phi$ only, such that we have
\begin{align}
	T(\phi)&=\left[\frac{45}{16\pi^2 g_*
          C}\phi^n\beta_{CI}^2(-W)^3\right]^{\frac{1}{4+m}}\;,
        \\ \mathcal{Q}(\phi)&=2C
        \left(\frac{45}{16\pi^2 g_*
          C}\right)^{\frac{m}{4+m}}\beta_{CI}^{\frac{2m}{4+m}}
        W^{\frac{2(m-2)}{4+m}}\phi^{-\frac{4n}{4+m}}\;.
\end{align}
Once again,  restricting to the chaotic class  gives
\begin{align}
	T(\phi)&= \left[\frac{45\alpha^2}{16\pi^2 g_*
          C} \left(-W_{\textit{f}}\right)^3 \left( \phi_{\textit{f}}^{-3\alpha/ 2} \right) \right]^{\frac{1}{4+m}}
        \phi^{\frac{3\alpha+2n-4}{2(4+m)}}\;,\\ \mathcal{Q}(\phi)&= 2C
        \left(\frac{45\alpha^2}{16\pi^2 g_*
          C}\right)^{\frac{m}{4+m}} \left( -W_\textit{f}\phi_\textit{f}^{- \alpha /2 } \right)^{\frac{2(m-2)}{4+m}} \phi^{\frac{\alpha(m-2)-2m-4n}{(4+m)}}\;.\label{eq:qofphistrongdisschaotic}
\end{align}
Recall that this regime is only attained for a particular set of
values for $\alpha$, $C$, $n$ and $m$ and therefore the consistency of
the condition $Q\gg1$ has to be checked explicitly model by model. The
dependence of $T$ and $Q$ on $\phi$ for different choices of $m$ and
$n$ are presented in Tab.~\ref{tab:PLBchaoticstrong}.
\begin{center}
\captionof{table}{Power-law behaviors of $\mathcal{Q}(\phi)$ and $T(\phi)$ for the chaotic class in the strong dissipative limit.}
\begin{tabular}{|l|c|c|}
	\hline
	\textbf{Dissipation Coefficient Ratio} & $\mathcal{Q}(\phi)\sim \phi ^ \#$ & $T(\phi)\sim \phi ^\#$\\
	\hline
	\text{Cubic ($m=3$, $n=2$) } & $ {\alpha}/{7}-2$ & ${{3\alpha}/{14}}$ \\
	\text{Linear ($m=1$, $n=0$) } & $ {-(\alpha+2)}/5$ & $(3\alpha-4)/10$ \\
  	\text{Inverse ($m=-1$, $n=0$) } & $ {(-3\alpha +2)}/3$ & $(3\alpha-4)/6$ \\
	\hline
\end{tabular}
\label{tab:PLBchaoticstrong}
\end{center}

Once again, a graphic representation of these behaviors can be seen in
{}Fig.~\ref{fig:asymptotic} shown in Sec.~\ref{sec:discussion_of_results}. Similar to the weak dissipation limit
case (in particular see footnote~\ref{foot:different_behaviors}),
different scalings can be obtained by considering different classes of
models, in particular choosing a different parameterization of
$\beta_{CI}$, which implies different expressions for $W(\phi)$. In
principle, by assuming that the strong dissipative regime holds during
the last $60$ e-foldings it could be possible to derive equations
similar to Eq.~\eqref{eq:N_of_phi_weak}, however, in most of the cases
this would not be physically relevant since we typically want $Q \ll
1$ at CMB scales.

\section{Discussion of the results}
\label{sec:discussion_of_results}
\begin{figure}[ht!]
    {\includegraphics[width=.475\columnwidth]{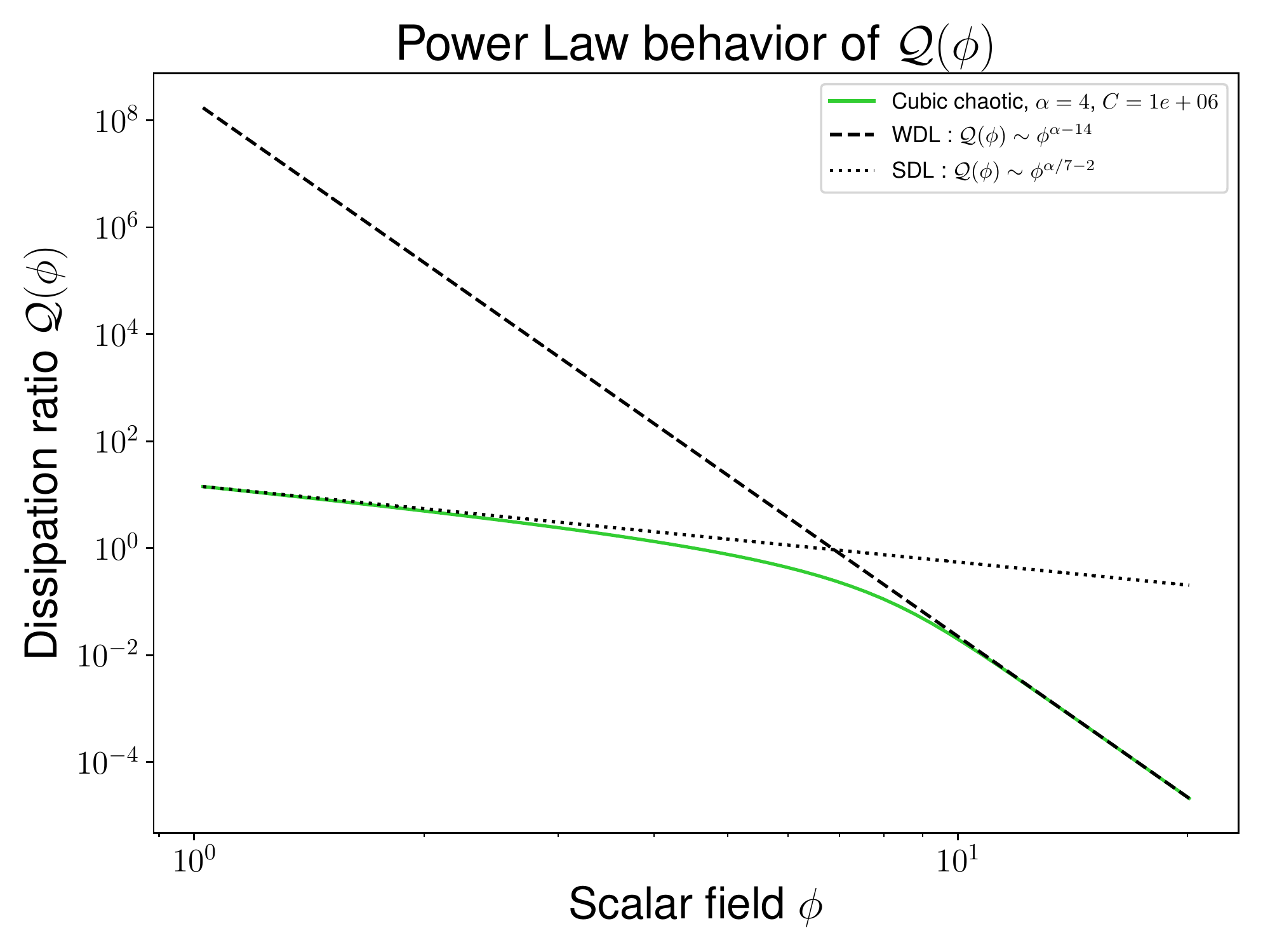}}
    \quad
        {\includegraphics[width=.475\columnwidth]{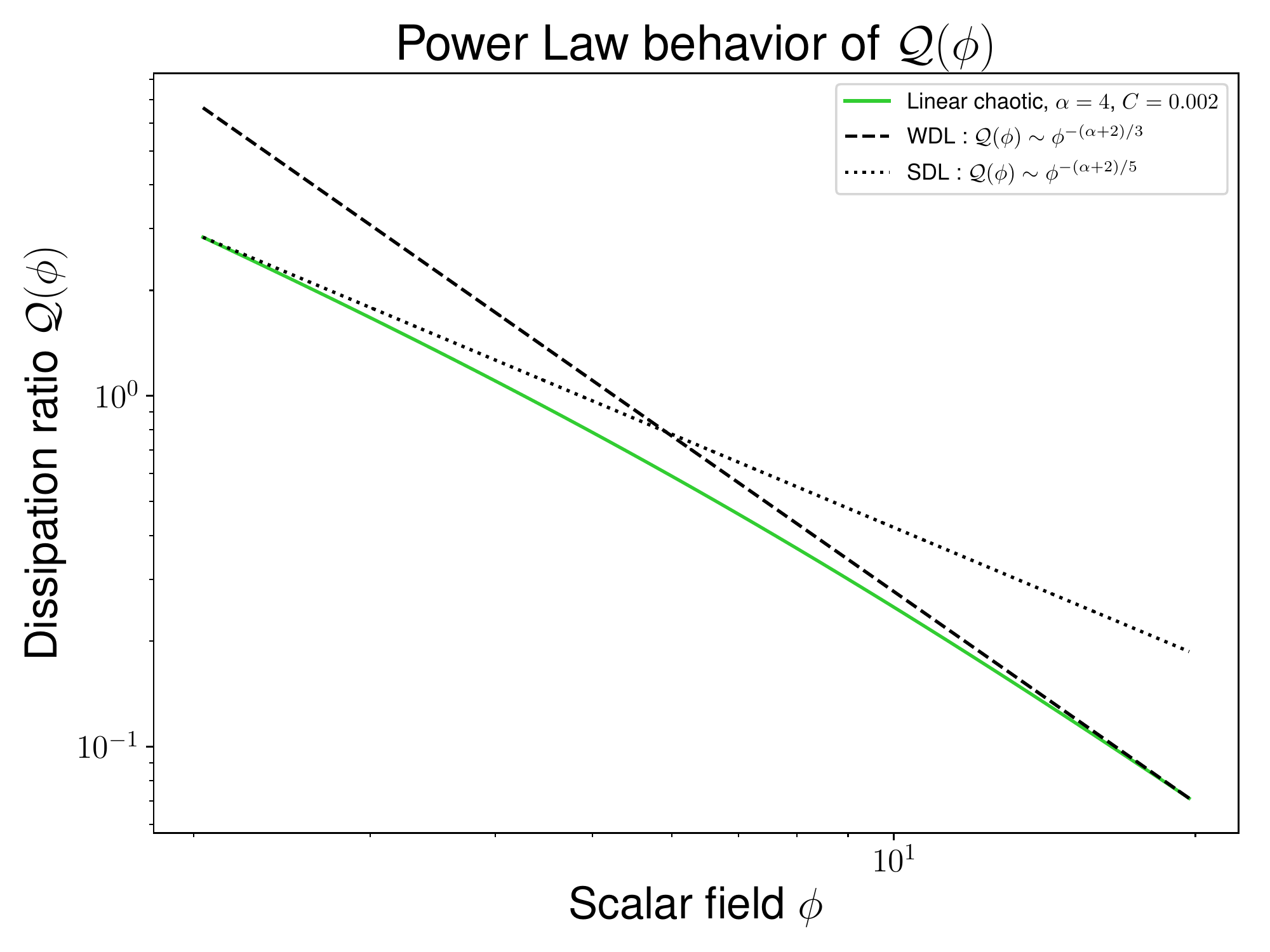}}\\ 
        {\includegraphics[width=.475\columnwidth]{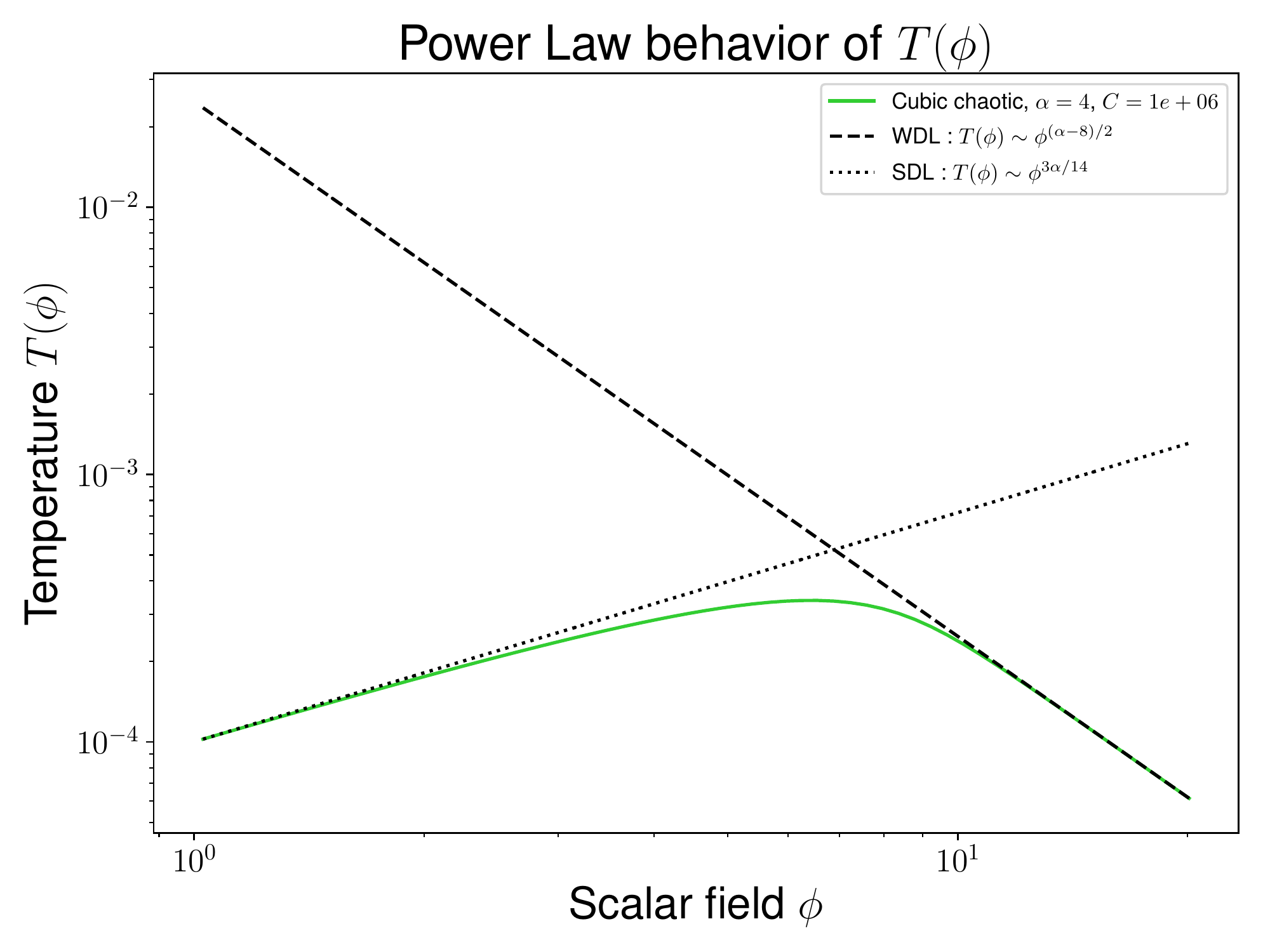}}
        \quad {
          \includegraphics[width=.475\columnwidth]{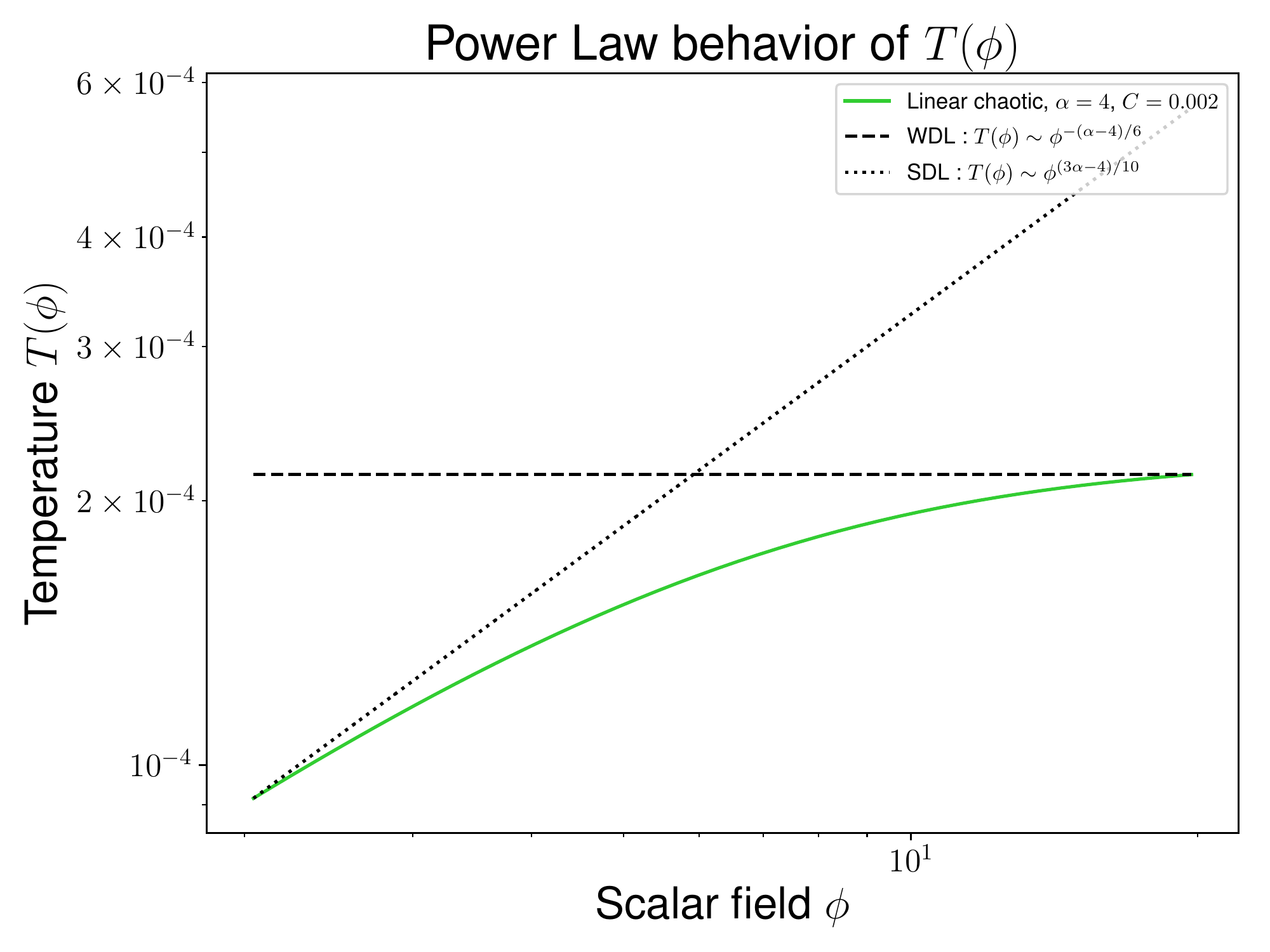}}
\caption{\label{fig:asymptotic} Loglog plots of the evolution of
  $\mathcal{Q}(\phi) \equiv Q(\phi,T(\phi))$ (top plots) and $T(\phi)$
  (bottom plots) during the last $60$ e-folds of inflation for $\alpha = 4$, $C = 10^6$ (left plots) with a
  dissipation coefficient ratio cubic in $T$ and $\alpha = 4$ and $C =
  2\times10^{-3}$ (right plots) with a dissipation coefficient ratio linear
  in $T$ compared with the analytical predictions of
  Sec.~\ref{sec:weak_strong_dissipation}. In these models $\phi$ decreases during inflation.}
\end{figure}

In this section we present and discuss the results of the numerical
analysis carried out by following the procedure outlined in
the previous section. While all the results shown are obtained by considering $\beta_{CI}$ of the chaotic
class, Eq.~\eqref{eq:chaoticbetaci}, a similar analysis can be performed for any other class\footnote{It is
  fair to stress that, according to the discussion of
  Sec.~\ref{sec:weak_strong_dissipation}, for different classes we
  expect qualitatively different results for the results shown in
  {}Fig.~\ref{fig:asymptotic}.},
such as the monomial ones, Eq.~\eqref{eq:monomial_class}, the inverse
type of potential, Eq.~\eqref{eq:inverse_class}, or the exponential
forms, Eq.~\eqref{eq:exponential_class}. Although we restrict to a
single choice for $\beta_{CI}$, we consider the four
different cases introduced in Sec.~\ref{sec:methods_and_results},
namely, the constant $Q = C$, cubic $Q = C T^3/(H \phi^2)$, linear $Q = C T/H $ and inverse $Q = C /( H T)$ forms of the dissipation coefficient ratio $Q$.

\begin{figure}[ht!]
    {\includegraphics[width=.475\columnwidth]{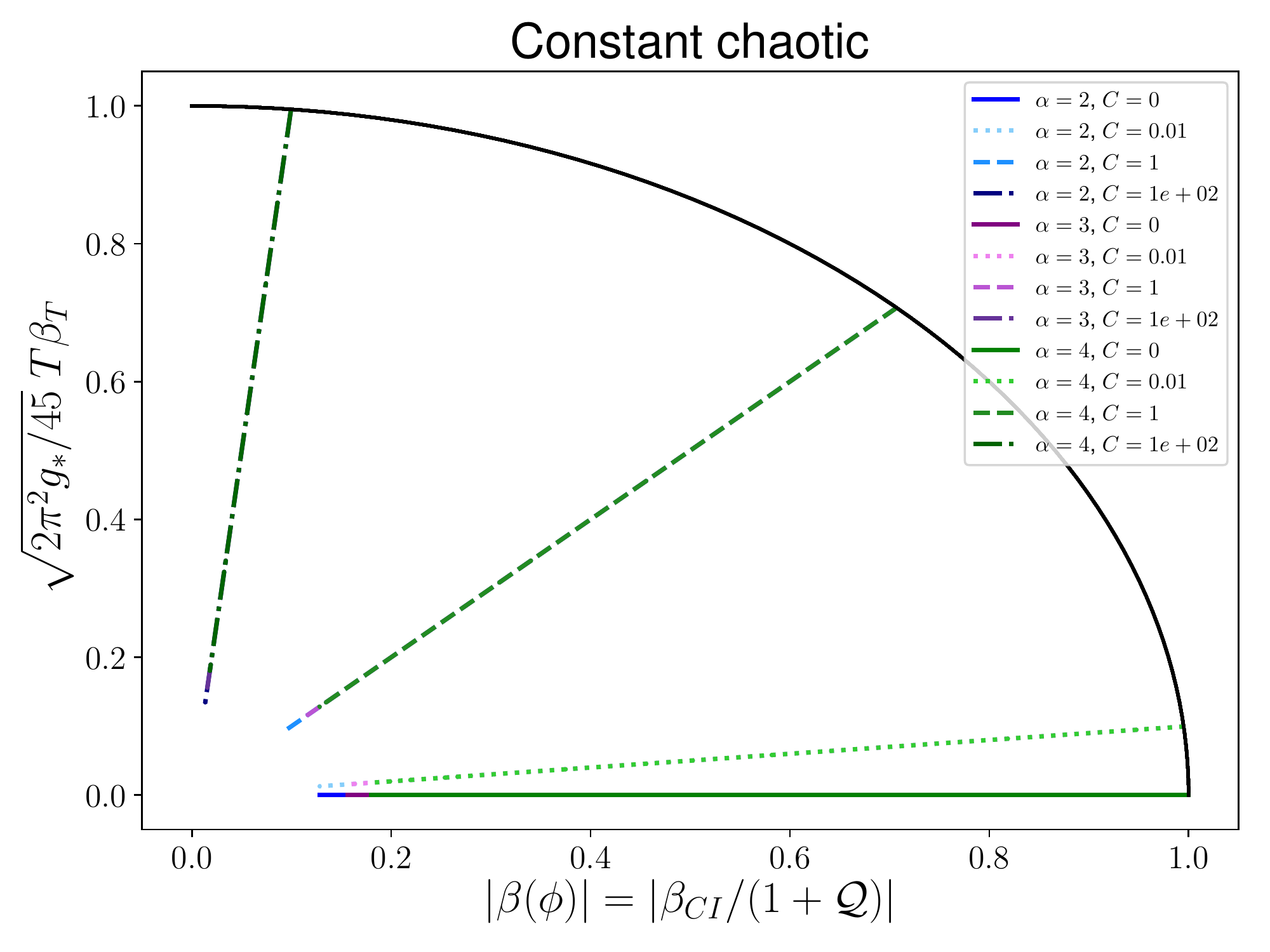}}
    \quad
        {\includegraphics[width=.475\columnwidth]{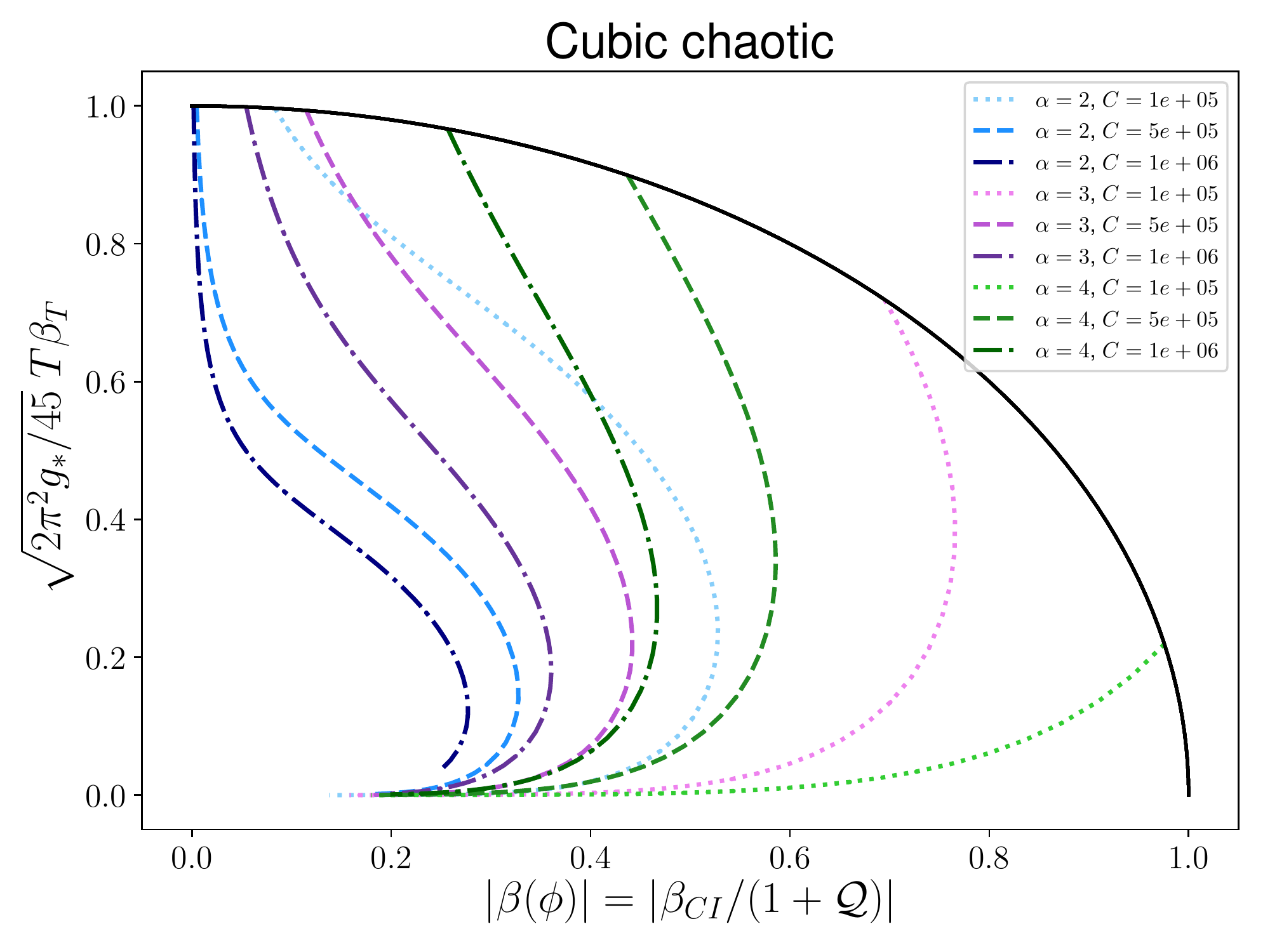}}
        \\ {\includegraphics[width=.475\columnwidth]{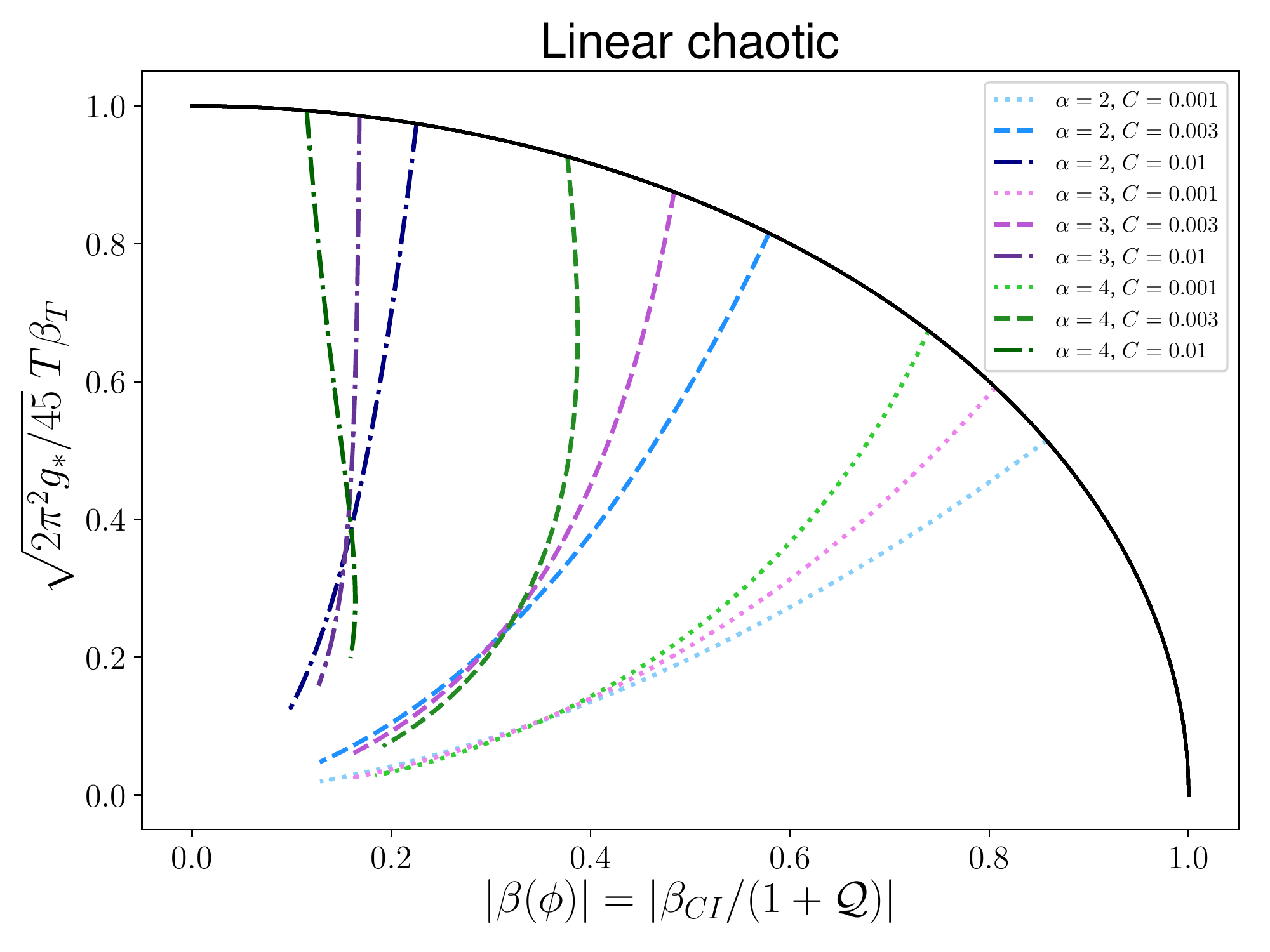}}
        \quad {
          \includegraphics[width=.475\columnwidth]{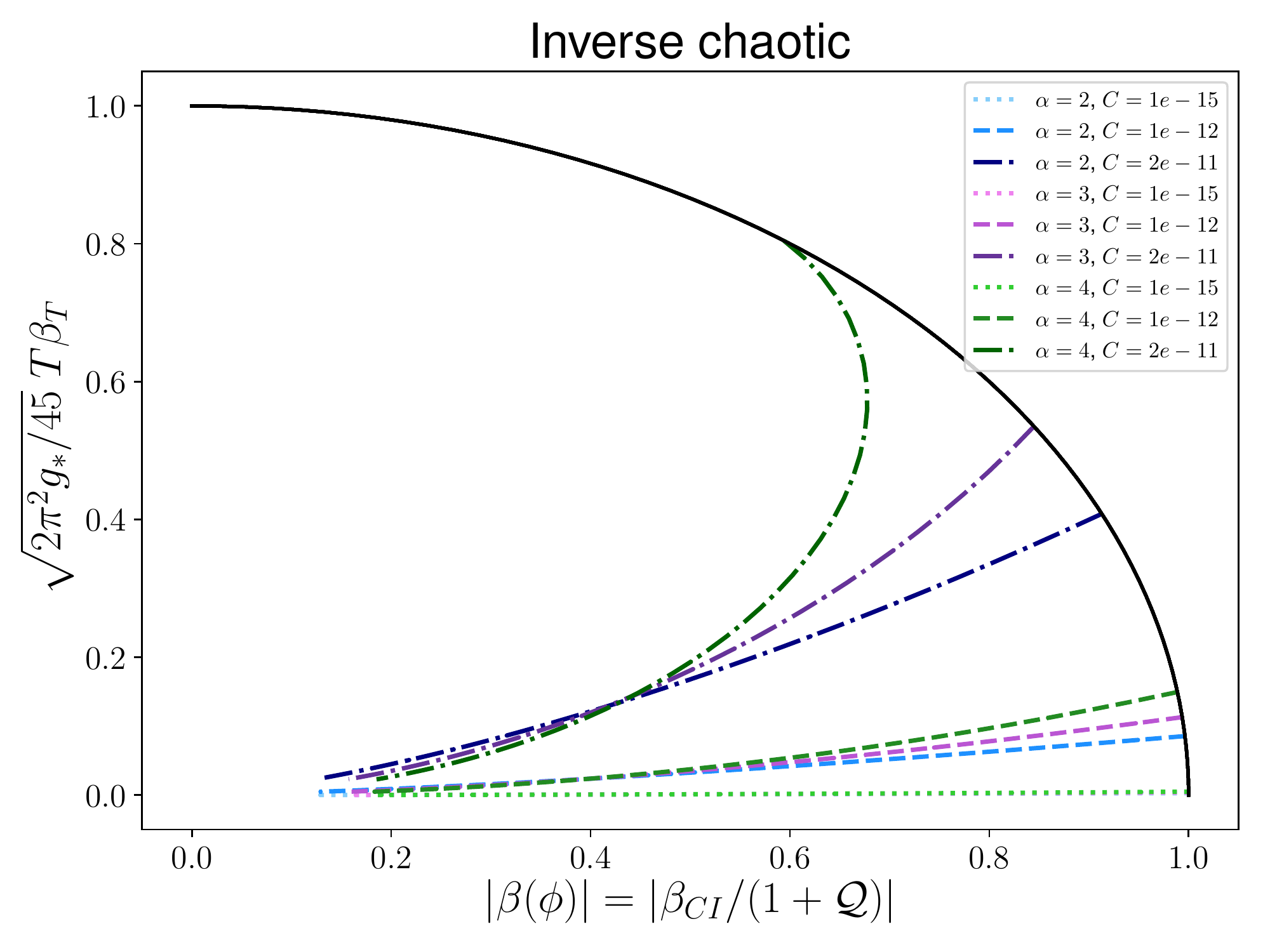}}
\caption{ \label{fig:2D_plots}2D plots to show the evolution of
  $\beta_{CI} $ and $T \beta_{T} $ for different dissipation
  coefficients. In particular, we show the evolution of models with
  (from top left to bottom right) $Q$ constant, cubic, linear and
  inversely proportional to $T$.}
\end{figure}

Let us start by discussing the evolution in the plane $(\beta_{CI}, T
\beta_{T} )$, shown in {}Fig.~\ref{fig:2D_plots}. The motivation for this kind of plots was explained in Sec.~\ref{sec:warm_beta}. As expected,
different parameterizations of the dissipation coefficient ratio lead to
different inflationary trajectories. Consistently with our
expectations, all the curves start from the neighborhood of the dS fixed point $(\beta_{CI},T \beta_{T}) = (0,0) $ and end onto the solid black curve, which represents the points in the
plane $(\beta_{CI}, T \beta_{T} )$ where inflation ends. One notes
that the straight trajectories of the constant case, among which we
have the standard cold case with $T \beta_T = 0$, are perfectly
consistent with the theoretical expectations; indeed from Eq.~\eqref{eq:Tbetaofphi} we see that $(T \beta_T)^2 \propto Q \beta_{CI}^2/(1 + Q)^2$. Interestingly, in many of these models inflation ends with
$\sqrt{2\pi^2g_* /45} \, T \beta_T = 4\rho_r/(9 H^2) \simeq 1$. This
implies that in these scenarios the amount of radiation present in the
Universe at the end of inflation is already sufficiently large to
quickly take over the inflaton energy density. As a consequence, 
already mentioned in Sec.~\ref{sec:warm_beta}, these models are not
expected to require an explosive (p)reheating to trigger the
transition from inflation to the radiation dominated phase.

Figure~\ref{fig:asymptotic} shows the evolution
of $T(\phi)$ and $\mathcal{Q}(\phi) \equiv Q(\phi,T(\phi))$ for two
illustrative cases, given by the values $\alpha = 2$ and $C = 10^5$,
with a dissipation coefficient ratio  cubic in $T$ and $\alpha = 4$, $C =
2\times10^{-3}$ with a dissipation coefficient ratio linear in $T$. During
inflation the field monotonically evolves from large to small values
and conversely the dissipation coefficient ratio (top panels of
{}Fig.~\ref{fig:asymptotic}) monotonically evolves from small to large
values. As a consequence, we expect the models to switch from the weak
dissipative regime $Q \ll 1$ to the strong dissipative limit $Q \gg 1$
 discussed in Sec.~\ref{sec:weak_strong_dissipation}.
 We expect the dissipation coefficient $\mathcal{Q}(\phi)$ to be monotonically growing with $\phi$ during the phase of inflation. 
 On the contrary, the radiation temperature $T(\phi)$ tends to approach the temperature of the ``thermal bath''
  of the inflaton energy density. As the latter is expected to
  slightly decrease during inflation, the expected behavior of radiation temperature $T(\phi)$ is to be decreasing towards the end of inflation after a possible initial phase of growth. The top and bottom panels of
{}Fig.~\ref{fig:asymptotic} clearly reproduce these behaviors for   $\mathcal{Q}(\phi)$ and $T(\phi)$.
 
 One notes that in both the plots of
$\mathcal{Q}$ and $T$, the curves are asymptotically approaching (both
for $Q \ll 1$ and for $Q \gg 1$) the power law behaviors predicted in
Sec.~\ref{sec:weak_strong_dissipation}. The transition from the weak
to the strong dissipative limit appears to be sharper in the Cubic
case. This is a direct consequence of the different dependences of
$\mathcal{Q}$ on $\phi$ in the asymptotic behaviors. Hence, the good
agreement between theory and numerical simulations confirms the
robustness of the numerical methods.

\begin{figure}[ht!]
    {\includegraphics[width=.475\columnwidth]{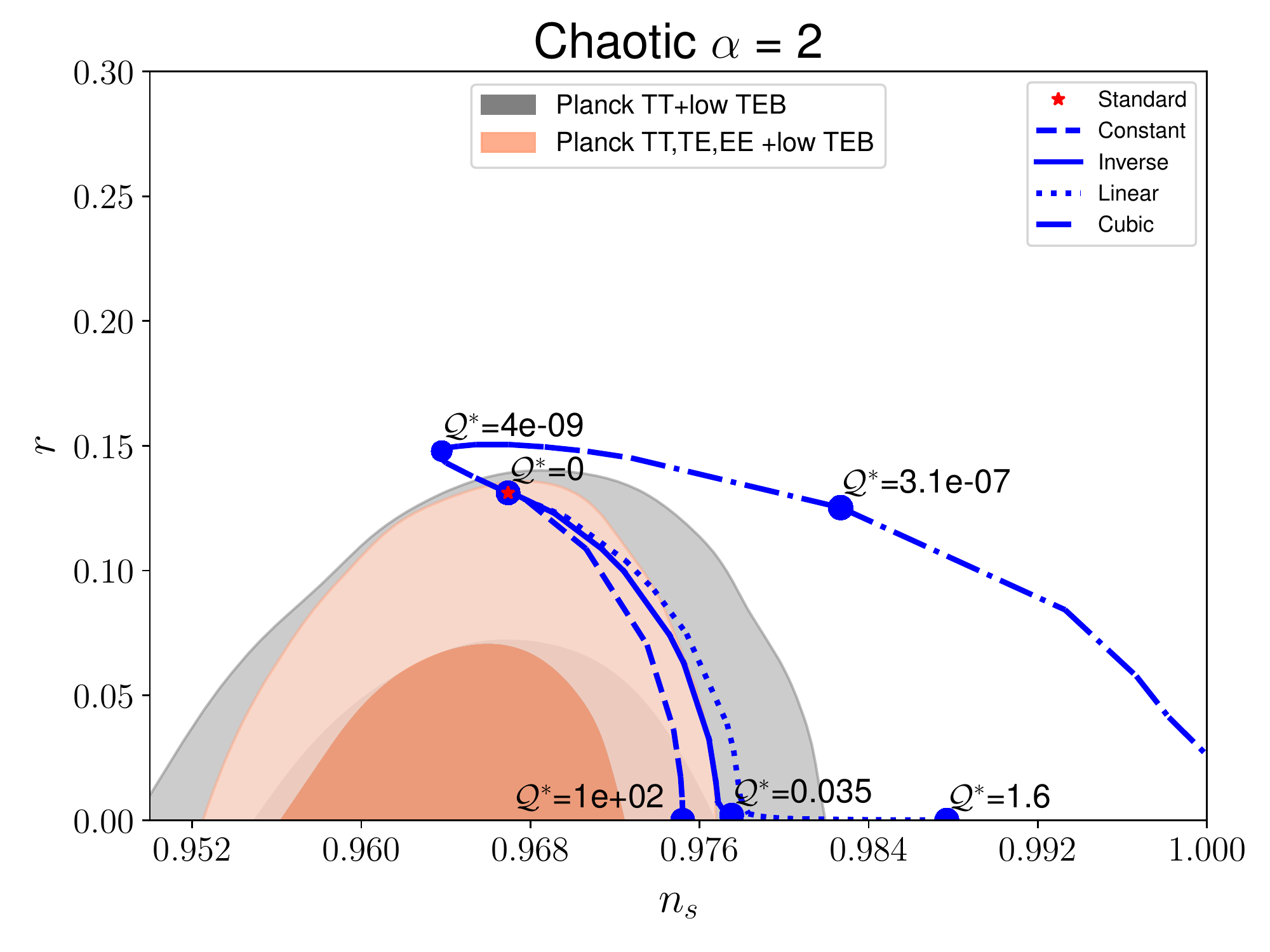}}
    \quad
        {\includegraphics[width=.475\columnwidth]{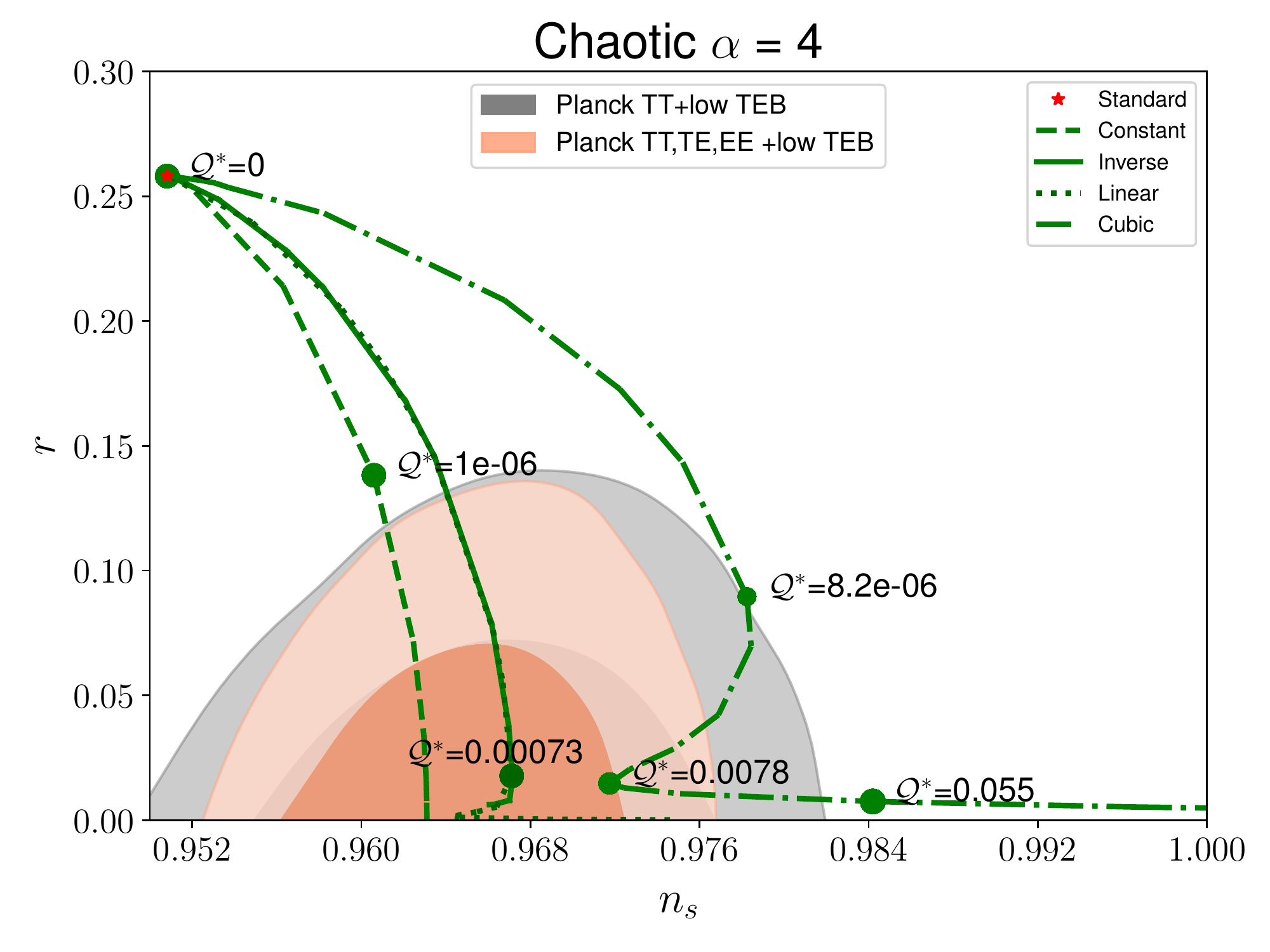}}
        \\ {\includegraphics[width=.475\columnwidth]{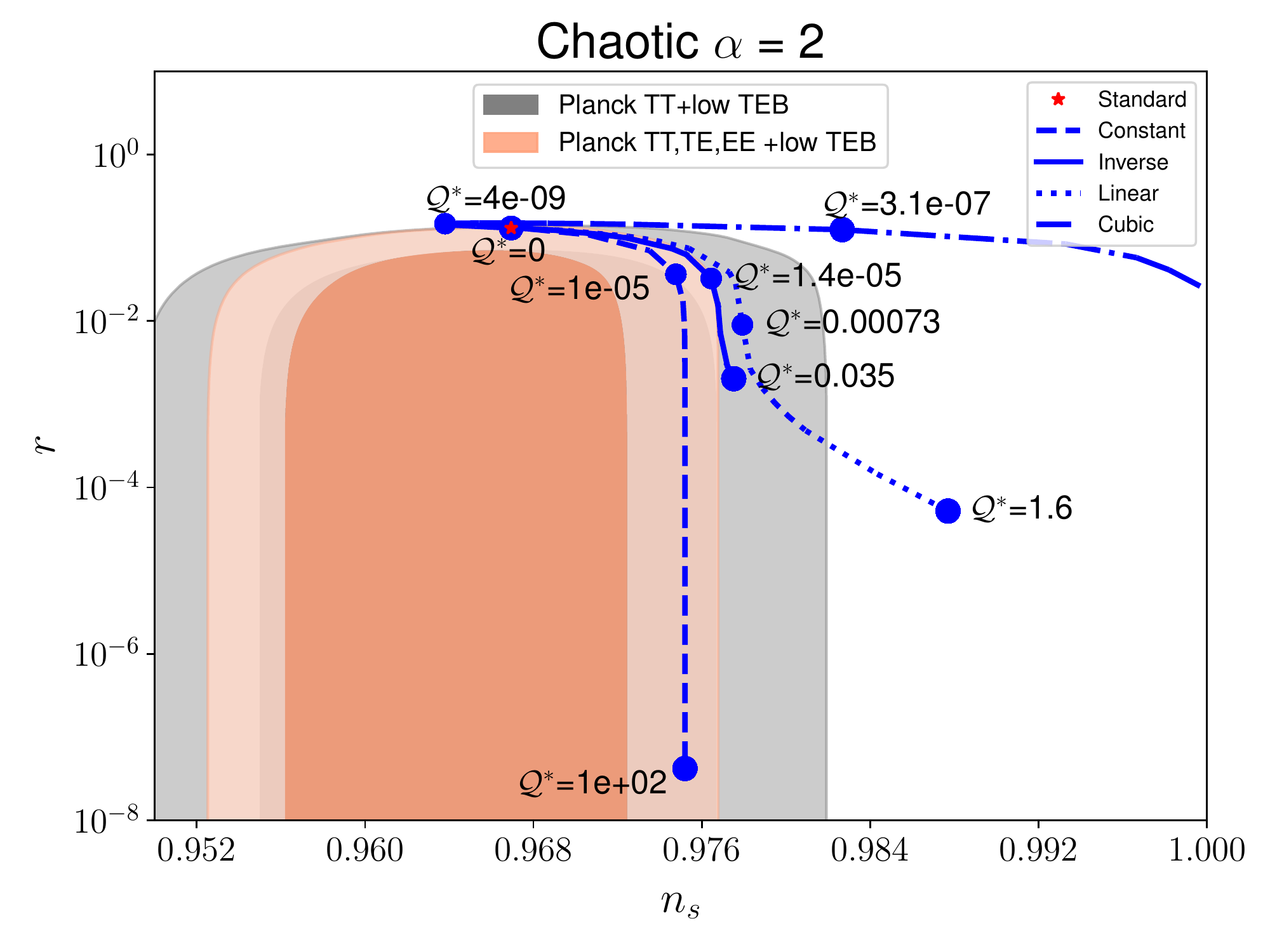}}
        \quad {
          \includegraphics[width=.475\columnwidth]{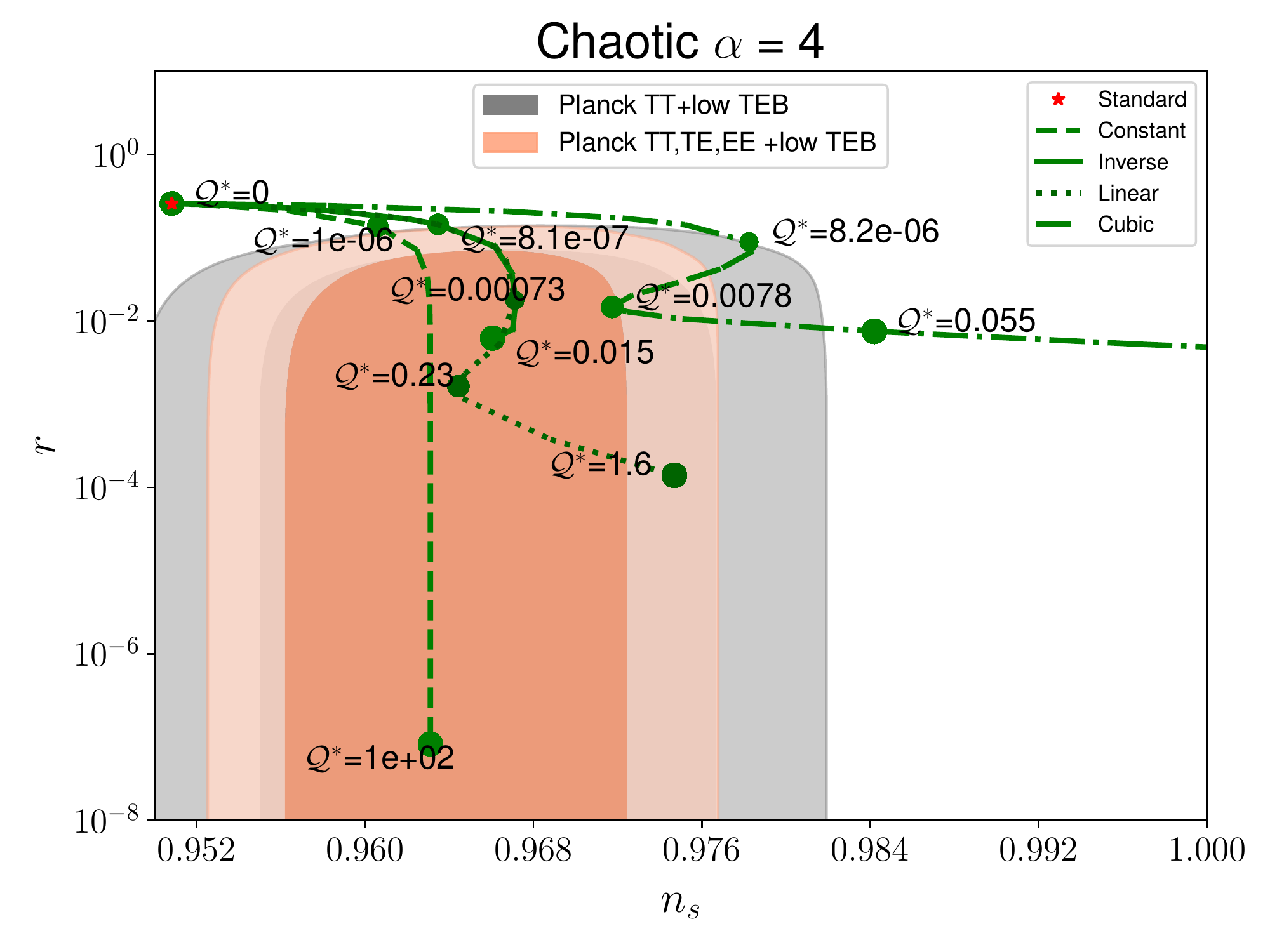}}
\caption{\label{fig:ns_r_evolution} Predictions for $n_s$ and $r $ in
  linear (top plots) and semilogarithmic (bottom plots) scale for a
  set of models of the chaotic class with $\alpha = 2 $ (left plots)
  and $\alpha = 4$ (right plots). For some of the models shown in this
  plot we report the value of $\mathcal{Q}$ at CMB scales (denoted
  with $\mathcal{Q}^*$).}
\end{figure}
 
For each model considered, a unique prediction for the scalar-spectral index and the tensor-to-scalar ratio is obtained. Figure~\ref{fig:ns_r_evolution} shows the
evolution of the predictions for $n_s$ and $r$ for  the chaotic class with different types of dissipation coefficient ratio. As expected, for very small values of
$\mathcal{Q}$ at CMB scales, the CMB observables
are, as expected, matching the predictions of the usual cold inflation
case, which in the plots shown in {}Fig.~\ref{fig:ns_r_evolution}, are
represented by a red star. {}For larger values of $\mathcal{Q}$ the predictions are modified as typically happens in warm
inflation. It is worth pointing out that the
modification of the predictions, see in particular the linear and
cubic cases with $\alpha = 4$, are qualitatively in agreement with the
results of Ref.~\cite{Benetti:2016jhf}. The small difference, at
around the $1\%$ level, in the predicted values of $n_s$  is mainly due to
slightly different values of $T$ in the numerical evolution and the
chosen value of $N_{CMB}$ used in the present work. As expected, the value of $n_s$
increases and the value of $r$ decreases with $\mathcal{Q}$ and $T$
and, thus, models which are in tension with (or even excluded by) the
Planck constraints in the cold case can be recovered in the warm
scenario. The sole exception to this behavior is the cubic case with
$\alpha = 2$ of Fig.~\ref{fig:ns_r_evolution}. In this case the values of $\mathcal{Q}$ and $T$ are
small at CMB scales implying that the spectrum is not modified by
thermal/dissipative effects. However, as at smaller scales the
production of radiation induces a friction that slows down the
evolution of the inflaton field, we see, similarly to
Ref.~\cite{Domcke:2016bkh}, a decrease of $n_s$ and an increase in $r$
due to shifting of the point of the potential probed by CMB
observables. Interestingly, in the quartic case ($\alpha=4$) the prediction for the inverse and the linear dissipation coefficient ratios are degenerate. This is explained by considering the field dependence of $\mathcal{Q}$ in the weak dissipative limit. From Tab.~\ref{tab:PLBchaoticweak} we read that in this regime both $\mathcal{Q}_{\text{inverse}}$ and $\mathcal{Q}_{\text{linear}}$ are proportional to $\phi^{-2}$ hence the similarity in the predictions.

\begin{figure}[ht!]
    {\includegraphics[width=.5\columnwidth]{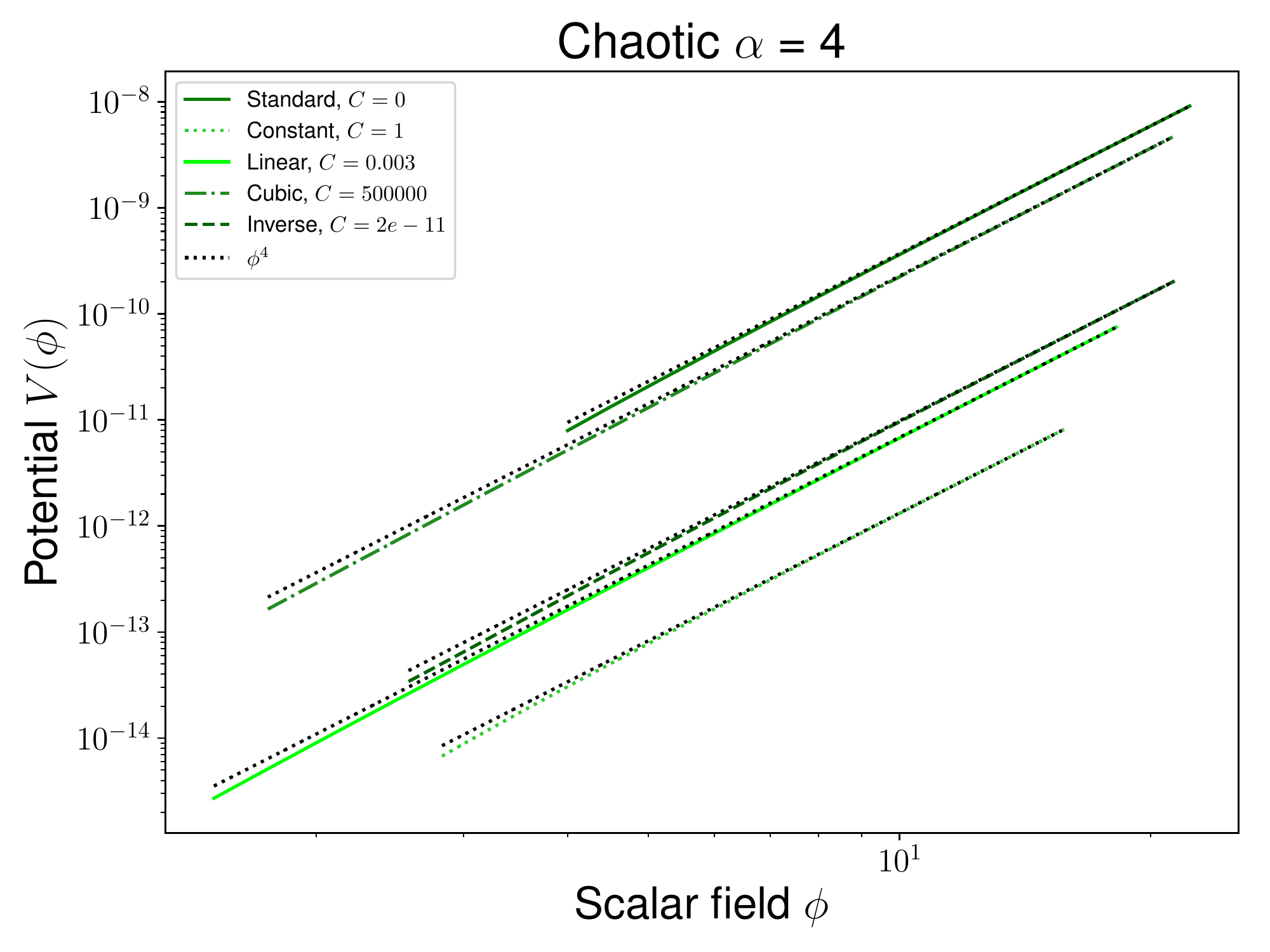}}
\caption{\label{fig:potential} Comparison between the inflationary
  potentials set by Eq.~\eqref{eq:potentialfrombeta} (for $\alpha =
  4$) and power laws potentials $V(\phi) = V_0 \phi^4$.}
\end{figure}

We conclude this section by presenting in {}Fig.~\ref{fig:potential} a
comparison between the inflationary potentials (calculated using
Eq.~\eqref{eq:potentialfrombeta}) for $\alpha = 4$ corresponding to
some of the cases discussed in this work and some power potentials of the
form
\begin{equation}
 	V(\phi) = V_0 \phi^\alpha \; .
 \end{equation} 
Note that the amplitudes are
always fixed in order to respect the COBE normalization. 
For $\alpha=2$ the $\phi$ dependence is the same
as in the well known case of chaotic inflation~\cite{Linde:1983gd}. As
expected, the two sets of curves are perfectly matching for large
values of $\phi$, meaning deep in the inflationary phase, where
$\beta_{CI}$ is much smaller than one and the potentials predicted by
Eq.~\eqref{eq:potentialfrombeta} are well approximated by power
laws. Conversely, for small values of $\phi$, higher order corrections
induce a deviation in $V(\phi)$ from the power law behavior observed
at large scales.  This type of analysis might be of particular use in
the problem of reconstructing potentials in warm
inflation~\cite{Herrera:2018cgi}.

\section{Conclusions and future perspectives}
\label{sec:conclusions_and_future}
In this work we have discussed the application of the $\beta$-function
formalism for inflation
to the case of warm inflation. We have shown in
Sec.~\ref{sec:warm_beta} that a consistent treatment of warm inflation
can be carried out in the language of this $\beta$-function
formalism. Interestingly, we have found that despite the presence of an additional functional freedom with respect to the cold case, a universal description still exists. For example, we have demonstrated that models with different functional forms for the dissipation coefficient ratios can give rise to very similar cosmological observables. Moreover, we have shown
that this formalism naturally offers an interesting graphical
representation of the inflationary phase in terms of bidimensional
plots in a plane of the variables $(\beta_{CI}, T \beta_{T} )$,
depicting the departure from the usual cold inflation case. A peculiar
property of these results is that they provide a clear insight on
the Universe energy budget in the last part of inflation, which in
turns allows us to infer some of the necessary properties of
(p)reheating.

We have also discussed in Sec.~\ref{sec:methods_and_results} the
definition of both numerical and analytical techniques used to perform
a systematic study of warm inflation within this framework. The
results of the numerical analysis were then presented and discussed in
Sec.~\ref{sec:discussion_of_results}. All the plots show an extremely good agreement between numerical
results and theoretical predictions. In particular, we stress the accuracy of the
predictions for the power law behaviors of the dissipation ratio $Q$
and temperature $T$ in both the small and large $Q$ limits. These analytical approximations could provide an extremely
useful tool for further studies on the topic. {}For example, by
studying the consistency of the conditions $Q \ll 1$ and $Q \gg 1$
with the analytical expressions, it is possible to understand at a
fully analytical level whether a given model could or could not access the
cold or warm regime respectively.

While in this paper our interest was mainly focused on the chaotic
class of potential, the generalization of the analysis to different
classes would be an interesting subject for future works on this topic
and should follow similarly the steps put forward in this work. In
particular, as already explained in
Sec.~\ref{sec:methods_and_results}, different scaling solutions (for
small and large $Q$) are expected to be obtained for different
classes. These analyses would be extremely useful in expanding and
strengthening our understanding of warm inflation. Moreover, the
deepening of our comprehension on the effects of interactions between
the inflaton and radiation could result in a definite step towards the
formulation of a theory of inflation which is somehow connected with
the rest of the fundamental interactions.

{}Finally, it is worth mentioning that in order to keep a direct
connection with previous works on this topic (and also with theory),
we always proceeded by first specifying $\beta_{CI}$ and $Q(\phi,T)$
and then computing $T(\phi)$ (and thus $\mathcal{Q}(\phi) \equiv
Q(\phi,T(\phi))$) by numerically solving
Eq.~\eqref{eq:Tempofphi}. However, it could also be equivalently
possible to start by fixing $\mathcal{Q}(\phi)$ and then identifying
the parameterizations of $Q(\phi,T)$ which correspond to this
choice. While formally these two possibilities are exactly equivalent,
the latter presents some computational advantages and has theoretical
interest, namely 
\begin{itemize}
	\item By starting with a fixed parameterization for
          $\mathcal{Q}(\phi)$ it could be possible to solve
          Eq.~\eqref{eq:Tempofphi} analytically. This implies that it
          could be possible to provide a full analytical treatment of
          some models of warm inflation;
	\item As a single parameterization of $\mathcal{Q}(\phi)$
          corresponds to several parameterizations of $Q(\phi,T)$, by
          specifying $\mathcal{Q}(\phi)$ we are not restricting our
          analysis to a single model but rather to a class of models
          sharing the same properties.  In this sense such an analysis
          would be more general than the one obtained by specifying
          $Q(\phi,T)$. Interestingly, the universality
          which is manifest at the background level is not expected to
          be broken by quantum perturbations. In particular this can
          be directly seen from Eq.~\eqref{eq:scalarspectrumbeta}-\eqref{eq:tensorspectrumbeta}, where it is manifest that
          all the quantities appearing in the expressions of the
          spectra can be directly computed once $\beta_{CI}$ and
          $\mathcal{Q}$ are specified.
\end{itemize}
Such an analysis would be an extremely interesting topic for future
studies on warm inflation. In particular, it would be
interesting to understand how, given a parameterization of
$\beta_{CI}$, it could be possible to reproduce the usual
parameterizations of $Q$ given, e.g., by
Eq.~\eqref{eq:Q_general_parameterization}, using $\mathcal{Q}(\phi)$.


\acknowledgements
A.B. is supported by STFC. J.M. is supported by Principals Career Development Scholarship and Edinburgh Global Research Scholarship.
M.P. acknowledges the support of the Spanish MINECOs ``Centro de
Excelencia Severo Ocho'' Programme under grant SEV-2012-0249. This
project has received funding from the European Unions Horizon 2020
research and innovation programme under the Marie Sk\l{}odowska-Curie
grant agreement No 713366.  R.\,O.\,R.~is partially supported by
Conselho Nacional de Desenvolvimento Cient\'{\i}fico e Tecnol\'ogico -
CNPq (Grant No. 302545/2017-4)   and Funda\c{c}\~ao Carlos Chagas Filho
de Amparo \`a Pesquisa do Estado do Rio de Janeiro - FAPERJ (Grant
No.~E-26/201.424/2014). The authors also acknowledge the kind hospitality and support
of the Higgs Centre for Theoretical Physics at the University of Edinburgh.

\appendix

\section{Compendium of useful formulae}
\label{sec:appendix_formula}
In this appendix we give some of the formula necessary to connect the analyses carried out in the framework of the
$\beta$-function formalism for inflation with the rest of the
literature on (warm) inflation, which is typically expressed in
terms of usual techniques relying on the specification of the
inflationary potential. As already discussed in
Sec.~\ref{sec:warm_beta}, any model of warm inflation can be
completely specified by a choice of $\beta_{CI}(\phi)$ and
$Q(T,\phi)$. The parameterizations of $\beta_{CI}$ are constrained by
the fact that inflation is realized by departing from a region with
$\beta_{CI} \ll 1$ and the parameterizations of $Q(T,\phi)$ are
typically fixed by theory. Once these two quantities are specified,
all the observables can be consistently expressed in term of
$\beta_{CI}(\phi)$ and $Q(T,\phi)$. It is important to stress that the
formalism not only applies to the background but also to
perturbations.

The Hubble slow-roll
parameters\footnote{Also known as Hubble flow functions
  (HFF)~\cite{Planck2015}.} $\epsilon_1\equiv - \dot{H} /H^2 $,
$\epsilon_{i+1}\equiv \dot{\epsilon}_i /(H\epsilon_{i})$ in terms of
$\beta_{CI}$ and $Q$, read
\begin{align}
  \epsilon_1&=\frac{1}{2}\frac{\beta_{CI}^2}{1+Q}=
  \frac{1}{2}(1+Q)\beta^2(\phi)\;,\\ \epsilon_2&
  =\frac{2\beta_{CI,\phi}}{1+Q}-\frac{\beta_{CI}Q_{,\phi}}{(1+Q)^2}\;,
  \\ \epsilon_3&=\frac{\beta_{CI}}{1+Q}
  \frac{\frac{2\beta_{CI,\phi\phi}}{1+Q}-\frac{3\beta_{CI,\phi}
      Q_{,\phi}}{(1+Q)^2}-\frac{\beta_{CI}Q_{,\phi\phi}}{(1+Q)^2}
    +\frac{2\beta_{CI}Q^2_{,\phi}}{(1+Q)^3}}{\frac{2\beta_{CI,\phi}}{1+Q}
    -\frac{\beta_{CI}Q_{,\phi}}{(1+Q)^2}}\;,\\
    \epsilon_{i+1}&=\frac{\beta_{CI}}{1+Q}\frac{\textrm{d}\ln\epsilon_i}{\textrm{d} \phi}\;.
\end{align} 
In order to have a better connection with the literature on warm
inflation, it is also useful to define 
\begin{align}
  \epsilon&\equiv\frac{1}{2}\frac{\beta_{CI}^2}{1+Q}\;,
  &&\eta\equiv\frac{2\beta_{CI,\phi}}{1+Q}\;,
  &&\sigma\equiv-\frac{\beta_{CI}Q_{,\phi}}{(1+Q)^2}\;,
\end{align}
 such that $\epsilon_1=\epsilon$ and $\epsilon_2=\eta+\sigma$.\\
 
Regarding cosmological perturbations, using the definitions
given in Sec.~\ref{sec:warm_beta} and in this Appendix we translate
the expressions of Sec.~\ref{sec:warm_perturbations} in terms of the
typical quantities of the $\beta$-function formalism. We start by
expressing the scalar spectrum in terms of $\beta_{CI}$, $\beta_T$ and
$\mathcal{Q}$,
\begin{align}
	\left.\Delta_s^2(k,\tau)\right|_{\tau=k^{-1}= -\frac{2}{a
            W}}&=\left[\frac{(1+Q)W}{4\pi\beta_{CI}}\right]^2
        \left(1+\frac{2}{\exp{\frac{1}{\beta_T}}-1}+\frac{\sqrt{12}\pi
          Q}{\sqrt{3+4\pi
            Q}}\beta_T\right)G(Q)\;.\label{eq:scalarspectrumbeta}
\end{align}
This expression
for the scalar spectrum is used to fix $W_{\textit{f}}$, the value of
the superpotential at the end of inflation, in order for the model to
agree with the COBE normalization~\cite{Planck2015,Ade:2015xua}. In
particular, we first derive $W(\phi)$ and then we impose $W_\textit{f}
= W(\phi(N_{CMB}))$ with $N_{CMB}=60$. {}For completeness, let us proceed by expressing the
tensor power spectrum as
\begin{equation}
	\left. \Delta^2_t (k)  \right|_{\tau=k^{-1}= -\frac{2}{a W}}
        =  \frac{ W^2}{2 \pi^2 }  \;, \label{eq:tensorspectrumbeta} 
\end{equation}
which has exactly the same expression as in the cold case. 

{}Finally, we provide the predictions for $n_s$ and $r$, with
expressions given by
\begin{align}
  n_s-1&=\frac{\beta_{CI}}{1+Q-\frac{1}{2}\beta_{CI}^2}
  \left[\frac{2Q_{,\phi}}{Q+1}-
  \beta_{CI}-\frac{2\beta_{CI,\phi}}{\beta_{CI}}+\frac{G_{,\phi}}{G}\right.\notag\\
  &\left.+ \frac{ 2n^2 e^{\frac{1}{\beta_T} } \frac{ \beta_{T,\phi} }{ \beta_{T}^2 } +
  \frac{ \sqrt{12} \pi \mathcal{Q}_{,\phi} }{ \sqrt{3 + 4 \pi
      \mathcal{Q} } } \beta_T -\frac{ \sqrt{3} \pi\mathcal{Q} }{
    \left(3 + 4 \pi \mathcal{Q}\right)^{3/2} } \left( 4 \pi
  \mathcal{Q}_{,\phi} \beta_T \right) +\frac{\sqrt{12} \pi
    \mathcal{Q}}{\sqrt{3 + 4 \pi \mathcal{Q}} } \beta_{T,\phi} } { 1
  +2n +\beta_{T} \frac{\sqrt{12}\pi Q }{ \sqrt{3 + 4 \pi \mathcal{Q}
}} } \right] \; ,\label{eq:nswarminfl}\\
  r&=\frac{8\beta_{CI}^2(\phi)}{(1+Q)^2}\frac{1}{\left(1+2n+\frac{2\sqrt{3}\pi
            Q}{\sqrt{3 + 4 \pi \mathcal{Q}}}
          \beta_{T}\right)G(Q)}\;.\label{eq:rwarminfl}
\end{align}
It has been a basic feature of the fluctuation-dissipation dynamics, intrinsic to warm inflation, that the tensor-to-scalar ratio in general is lower as compared to cold inflation. For the $\phi^4$ model, it was predicted from warm inflation in~\cite{Berera:1999ws,BasteroGil:2009ec}, well before the CMB data, that this ratio would be lower. Here we present a compact expression for the tensor-to-scalar ratio simply written  as the expression that appears for cold inflation $r_{CI}=8\beta_{CI}^2(\phi)$ multiplied by a correction factor, the denominator of Eq.~\eqref{eq:rwarminfl}. In agreement with the literature~\cite{Berera:1999ws,BasteroGil:2009ec} and as discussed in section~\ref{sec:warm_perturbations} this correction term lowers the prediction for $r$ with respect to cold inflation when the dissipation coefficient ratio is of order of unity or when $\beta_T$ is larger than one.

\subsection{Some additional universality classes}
\label{sec:universality_classes}

Let us conclude this Appendix by briefly presenting some of the
classes introduced in Ref.~\cite{Binetruy:2014zya}, starting with the
so-called {monomial class}, where
\begin{equation}
	\label{eq:monomial_class}
	\beta_{CI} = \alpha \phi^q \; ,
\end{equation}
with $\alpha$ and $q$ being positive constants. This class describes small field models, i.e., inflation take place for
$\phi \ll 1$, with
\begin{equation}
	\label{eq:monomial_superpot}
	W(\phi) = W_{\textit{f}} \exp \left[- \frac{\alpha}{2 (q+1)}
        \left(\phi^{q+1} - \phi_{\textit{f}}^{q+1}\right)
        \right] \; ,
\end{equation}
implying that at the lowest order models of this class feature a
hilltop potential. 

We can also consider the so-called {inverse class}, where
\begin{equation}
	\label{eq:inverse_class}
	\beta_{CI} = -\frac{\alpha}{\phi^q} \; ,
\end{equation}
with $\alpha$ and $q$ being positive constants. This class describes large field models, i.e., inflation take place for
$\phi \gg 1$, with
\begin{equation}
	\label{eq:inverse_superpot}
	W(\phi) =  W_{\textit{f}} \exp \left[ \frac{\alpha}{2 (q-1)}
        \left( \frac{1}{ \phi_{\textit{f}}^{q-1} } - \frac{1}{ \phi
            ^{q-1}} \right)   \right] \; ,
\end{equation}
implying that at the lowest order models of this class feature an
algebraically flat plateau potential. 

{}Finally, is the so-called {exponential class}, where
\begin{equation}
	\label{eq:exponential_class}
	\beta_{CI} = - \alpha \exp(- \gamma \phi) \; , 
\end{equation}
with $\alpha$ and $\gamma$ being positive constants. This class describes large field models, with
\begin{equation}
	\label{eq:exponential_superpot}
	W(\phi) = W_{\textit{f}} \exp \left\{ - \frac{\alpha}{2\gamma}
        \left[  \exp\left( -\gamma \phi \right) -\exp\left( -\gamma
          \phi_{\textit{f}} \right)  \right] \right\} \; ,
\end{equation}
implying that at the lowest order models of this class feature an
exponentially flat plateau potential.


\begin{thebibliography}{99}


\bibitem{Planck2015}   
P.~A.~R.~Ade {\it et al.} [Planck Collaboration],
\emph{Planck 2015 results. XX. Constraints on inflation},
  Astron.\ Astrophys.\  {\bf 594}, A20 (2016).
  [arXiv:1502.02114 [astro-ph.CO]].

\bibitem{Ade:2015ava}
  P.~A.~R.~Ade {\it et al.} [Planck Collaboration],
  ``Planck 2015 results. XVII. Constraints on primordial non-Gaussianity'',
  Astron.\ Astrophys.\  {\bf 594} (2016) A17
  [arXiv:1502.01592 [astro-ph.CO]].

\bibitem{Ade:2015xua}
  P.~A.~R.~Ade {\it et al.} [Planck Collaboration],
  ``Planck 2015 results. XIII. Cosmological parameters'',
  Astron.\ Astrophys.\  {\bf 594} (2016) A13
  [arXiv:1502.01589 [astro-ph.CO]].


\bibitem{Binetruy:2014zya}
  P.~Binetruy, E.~Kiritsis, J.~Mabillard, M.~Pieroni and C.~Rosset,
  ``Universality classes for models of inflation'',
  JCAP {\bf 1504} (2015) no.04,  033
  [arXiv:1407.0820 [astro-ph.CO]].

\bibitem{Pieroni:2015cma}
  M.~Pieroni,
  ``$\beta$-function formalism for inflationary models with a non minimal coupling with gravity'',
  JCAP {\bf 1602} (2016) no.02,  012
  [arXiv:1510.03691 [hep-ph]].
  
\bibitem{Pieroni:2016gdg}
  M.~Pieroni,
  ``Classification of inflationary models and constraints on fundamental physics'',
  PhD thesis, APC, Paris, 2016.
  [arXiv:1611.03732 [gr-qc]].


\bibitem{Binetruy:2016hna}
  P.~Binétruy, J.~Mabillard and M.~Pieroni,
  ``Universality in generalized models of inflation'',
  JCAP {\bf 1703} (2017) no.03,  060
  [arXiv:1611.07019 [gr-qc]].

\bibitem{Cicciarella:2016dnv}
  F.~Cicciarella and M.~Pieroni,
  ``Universality for quintessence'',
  JCAP {\bf 1708} (2017) no.08,  010
  [arXiv:1611.10074 [gr-qc]].

\bibitem{Cicciarella:2017nls}
  F.~Cicciarella, J.~Mabillard and M.~Pieroni,
  ``New perspectives on constant-roll inflation'',
  JCAP {\bf 1801} (2018) no.01,  024
  [arXiv:1709.03527 [astro-ph.CO]].

\bibitem{Salopek:1990jq}
  D.~S.~Salopek and J.~R.~Bond,
  ``Nonlinear evolution of long wavelength metric fluctuations in inflationary models'',
  Phys.\ Rev.\ D {\bf 42} (1990) 3936.

\bibitem{Skenderis:2006jq}
  K.~Skenderis and P.~K.~Townsend,
  ``Hidden supersymmetry of domain walls and cosmologies'',
  Phys.\ Rev.\ Lett.\  {\bf 96} (2006) 191301
  [arXiv:0602260 [hep-th]].

\bibitem{McFadden:2009fg}
  P.~McFadden and K.~Skenderis,
  ``Holography for Cosmology'',
  Phys.\ Rev.\ D {\bf 81} (2010) 021301
  [arXiv:0907.5542 [hep-th]].


\bibitem{McFadden:2010na}
  P.~McFadden and K.~Skenderis,
  ``The Holographic Universe'',
  J.\ Phys.\ Conf.\ Ser.\  {\bf 222} (2010) 012007
  [arXiv:1001.2007 [hep-th]].

\bibitem{Maldacena:1997re}
  J.~M.~Maldacena,
  ``The Large N limit of superconformal field theories and supergravity'',
  Int.\ J.\ Theor.\ Phys.\  {\bf 38} (1999) 1113
   [Adv.\ Theor.\ Math.\ Phys.\  {\bf 2} (1998) 231]
  [arXiv:9711200 [hep-th]].

\bibitem{McFadden:2010vh}
  P.~McFadden and K.~Skenderis,
  ``Holographic Non-Gaussianity'',
  JCAP {\bf 1105} (2011) 013
  [arXiv:1011.0452 [hep-th]].

\bibitem{Bzowski:2012ih}
  A.~Bzowski, P.~McFadden and K.~Skenderis,
  ``Holography for inflation using conformal perturbation theory'',
  JHEP {\bf 1304} (2013) 047
  [arXiv:1211.4550 [hep-th]].

\bibitem{Garriga:2014ema}
  J.~Garriga and Y.~Urakawa,
  ``Holographic inflation and the conservation of $\zeta$'',
  JHEP {\bf 1406} (2014) 086
  [arXiv:1403.5497 [hep-th]].

\bibitem{Garriga:2014fda}
  J.~Garriga, K.~Skenderis and Y.~Urakawa,
  ``Multi-field inflation from holography'',
  JCAP {\bf 1501} (2015) no.01,  028
  [arXiv:1410.3290 [hep-th]].

\bibitem{Afshordi:2016dvb}
  N.~Afshordi, C.~Coriano, L.~Delle Rose, E.~Gould and K.~Skenderis,
  ``From Planck data to Planck era: Observational tests of Holographic Cosmology'',
  Phys.\ Rev.\ Lett.\  {\bf 118} (2017) no.4,  041301
  [arXiv:1607.04878 [astro-ph.CO]].

\bibitem{Afshordi:2017ihr}
  N.~Afshordi, E.~Gould and K.~Skenderis,
  ``Constraining holographic cosmology using Planck data'',
  Phys.\ Rev.\ D {\bf 95} (2017) no.12,  123505
  [arXiv:1703.05385 [astro-ph.CO]].

\bibitem{Hawking:2017wrd}
  S.~W.~Hawking and T.~Hertog,
  ``A Smooth Exit from Eternal Inflation'',
  [arXiv:1707.07702 [hep-th]].

\bibitem{Conti:2017pqc}
  G.~Conti, T.~Hertog and Y.~Vreys,
  ``Holographic Measure on Eternal Inflation'',
  [arXiv:1707.09663 [hep-th]].

\bibitem{Berera:1995wh}
  A.~Berera and L.~Z.~Fang,
  ``Thermally induced density perturbations in the inflation era'',
  Phys.\ Rev.\ Lett.\  {\bf 74} (1995) 1912
  [arXiv:9501024 [astro-ph]].
  
\bibitem{Berera:1995ie}
  A.~Berera,
  ``Warm inflation'',
  Phys.\ Rev.\ Lett.\  {\bf 75} (1995) 3218
  [arXiv:9509049 [astro-ph]].
  
\bibitem{Berera:1998px}
  A.~Berera, M.~Gleiser and R.~O.~Ramos,
  ``A First principles warm inflation model that solves the cosmological horizon / flatness problems'',
  Phys.\ Rev.\ Lett.\  {\bf 83} (1999) 264
  [arXiv:9809583 [hep-ph]].

\bibitem{Berera:2002sp}
  A.~Berera and R.~O.~Ramos,
  ``Construction of a robust warm inflation mechanism'',
  Phys.\ Lett.\ B {\bf 567} (2003) 294
  [arXiv:0210301 [hep-ph]].


\bibitem{Bastero-Gil:2016qru}
  M.~Bastero-Gil, A.~Berera, R.~O.~Ramos and J.~G.~Rosa,
  ``Warm Little Inflaton'',
  Phys.\ Rev.\ Lett.\  {\bf 117} (2016) no.15,  151301
  [arXiv:1604.08838 [hep-ph]].

\bibitem{Berera:2008ar}
  A.~Berera, I.~G.~Moss and R.~O.~Ramos,
  ``Warm Inflation and its Microphysical Basis'',
  Rept.\ Prog.\ Phys.\  {\bf 72} (2009) 026901
  [arXiv:0808.1855 [hep-ph]].

\bibitem{Ramos:2013nsa}
  R.~O.~Ramos and L.~A.~da Silva,
  ``Power spectrum for inflation models with quantum and thermal noises'',
  JCAP {\bf 1303} (2013) 032
  [arXiv:1302.3544 [astro-ph.CO]].

\bibitem{Bartrum:2013fia}
  S.~Bartrum, M.~Bastero-Gil, A.~Berera, R.~Cerezo, R.~O.~Ramos and J.~G.~Rosa,
  ``The importance of being warm (during inflation)'',
  Phys.\ Lett.\ B {\bf 732} (2014) 116
  [arXiv:1307.5868 [hep-ph]].
  
\bibitem{Bastero-Gil:2014jsa}
  M.~Bastero-Gil, A.~Berera, I.~G.~Moss and R.~O.~Ramos,
  ``Cosmological fluctuations of a random field and radiation fluid'',
  JCAP {\bf 1405} (2014) 004
  [arXiv:1401.1149 [astro-ph.CO]].


\bibitem{Bastero-Gil:2014raa}
  M.~Bastero-Gil, A.~Berera, I.~G.~Moss and R.~O.~Ramos,
  ``Theory of non-Gaussianity in warm inflation'',
  JCAP {\bf 1412} (2014) no.12,  008
  [arXiv:1408.4391 [astro-ph.CO]].

\bibitem{Moss:2007cv}
  I.~G.~Moss and C.~Xiong,
  ``Non-Gaussianity in fluctuations from warm inflation'',
  JCAP {\bf 0704} (2007) 007
  [arxiv:0701302 [astro-ph]].

\bibitem{Vicente:2015hga}
  G.~S.~Vicente, L.~A.~da Silva and R.~O.~Ramos,
  ``Eternal inflation in a dissipative and radiation environment: Heated demise of eternity'',
  Phys.\ Rev.\ D {\bf 93} (2016) no.6,  063509
  [arXiv:1509.08983 [astro-ph.CO]].
  
  
\bibitem{Arya:2017zlb}
  R.~Arya, A.~Dasgupta, G.~Goswami, J.~Prasad and R.~Rangarajan,
  ``Revisiting CMB constraints on warm inflation'',
  JCAP {\bf 1802} (2018) no.02,  043
  [arXiv:1710.11109 [astro-ph.CO]].
  
\bibitem{Bastero-Gil:2017wwl}
  M.~Bastero-Gil, S.~Bhattacharya, K.~Dutta and M.~R.~Gangopadhyay,
  ``Constraining Warm Inflation with CMB data'',
  JCAP {\bf 1802} (2018) no.02,  054
  [arXiv:1710.10008 [astro-ph.CO]].

\bibitem{Rangarajan:2018tte}
  R.~Rangarajan,
  ``Current Status of Warm Inflation'',
  [arXiv:1801.02648 [astro-ph.CO]].
 

\bibitem{Schwinger:1951nm}
  J.~S.~Schwinger,
  ``On gauge invariance and vacuum polarization'',
  Phys.\ Rev.\  {\bf 82} (1951) 664.

\bibitem{Hayashinaka:2016qqn}
  T.~Hayashinaka, T.~Fujita and J.~Yokoyama,
  ``Fermionic Schwinger effect and induced current in de Sitter space'',
  JCAP {\bf 1607} (2016) no.07,  010
  [arXiv:1603.04165 [hep-th]].

\bibitem{Tangarife:2017rgl}
  W.~Tangarife, K.~Tobioka, L.~Ubaldi and T.~Volansky,
  ``Dynamics of Relaxed Inflation'',
  JHEP {\bf 1802} (2018) 084
  [arXiv:1706.03072 [hep-ph]].

\bibitem{Ferreira:2017lnd}
  R.~Z.~Ferreira and A.~Notari,
  ``Thermalized Axion Inflation'',
  JCAP {\bf 1709} (2017) no.09,  007
  [arXiv:1706.00373 [astro-ph.CO]].
  

\bibitem{Graham:2008vu}
  I.~G.~Moss and C.~M.~Graham,
  ``Particle production and reheating in the inflationary universe'',
  Phys.\ Rev.\ D {\bf 78} (2008) 123526
  [arXiv:0810.2039 [hep-ph]].


\bibitem{BasteroGil:2010pb}
  M.~Bastero-Gil, A.~Berera and R.~O.~Ramos,
  ``Dissipation coefficients from scalar and fermion quantum field interactions'',
  JCAP {\bf 1109} (2011) 033
  [arXiv:1008.1929 [hep-ph]].
  
\bibitem{BasteroGil:2012cm}
  M.~Bastero-Gil, A.~Berera, R.~O.~Ramos and J.~G.~Rosa,
  ``General dissipation coefficient in low-temperature warm inflation'',
  JCAP {\bf 1301} (2013) 016
  [arXiv:1207.0445 [hep-ph]].


\bibitem{Berera:1999ws}
  A.~Berera,
  ``Warm inflation at arbitrary adiabaticity: A Model, an existence proof for inflationary dynamics in quantum field theory'',
  Nucl.\ Phys.\ B {\bf 585} (2000) 666
  [hep-ph/9904409].


\bibitem{Berera:2004vm}
  A.~Berera,
  ``Warm inflation solution to the eta problem'',
  PoS AHEP {\bf 2003} (2003) 069
  [arXiv:0401139 [hep-ph]].
  
  
\bibitem{Graham:2009bf}
  C.~Graham and I.~G.~Moss,
  ``Density fluctuations from warm inflation'',
  JCAP {\bf 0907} (2009) 013
  [arXiv:0905.3500 [astro-ph.CO]].
  
\bibitem{BasteroGil:2011xd}
  M.~Bastero-Gil, A.~Berera and R.~O.~Ramos,
  ``Shear viscous effects on the primordial power spectrum from warm inflation'',
  JCAP {\bf 1107} (2011) 030
  [arXiv:1106.0701 [astro-ph.CO]].
  
\bibitem{BasteroGil:2009ec}
  M.~Bastero-Gil and A.~Berera,
  ``Warm inflation model building,''
  Int.\ J.\ Mod.\ Phys.\ A {\bf 24} (2009) 2207
  [arXiv:0902.0521 [hep-ph]].

 
\bibitem{Taylor:2000ze}
  A.~N.~Taylor and A.~Berera,
  ``Perturbation spectra in the warm inflationary scenario'',
  Phys.\ Rev.\ D {\bf 62} (2000) 083517
  [arXiv:0006077 [astro-ph]].


\bibitem{Benetti:2016jhf}
  M.~Benetti and R.~O.~Ramos,
  ``Warm inflation dissipative effects: predictions and constraints from the Planck data'',
  Phys.\ Rev.\ D {\bf 95} (2017) no.2,  023517
  [arXiv:1610.08758 [astro-ph.CO]].
  
  


\bibitem{Bastero-Gil:2017yzb}
  M.~Bastero-Gil, A.~Berera, R.~O.~Ramos and J.~G.~Rosa,
  ``Adiabatic out-of-equilibrium solutions to the Boltzmann equation in warm inflation'',
  JHEP {\bf 1802} (2018) 063
  [arXiv:1711.09023 [hep-ph]].
  
\bibitem{Domcke:2016bkh}
  V.~Domcke, M.~Pieroni and P.~Binétruy,
  ``Primordial gravitational waves for universality classes of pseudoscalar inflation'',
  JCAP {\bf 1606} (2016) 031
  [arXiv:1603.01287 [astro-ph.CO]].

\bibitem{Binetruy:2006ad}
  P.~Binetruy,
  ``Supersymmetry: Theory, experiment and cosmology'',
  Oxford, UK: Oxford Univ. Pr. (2006) 520 p

\bibitem{Zhang:2009ge}
Y.~Zhang,
``Warm Inflation With A General Form Of The Dissipative Coefficient'',
JCAP {\bf 0903} (2009) 023
[arXiv:0903.0685 [hep-ph]].


  
\bibitem{Sayar:2017pam}
  K.~Sayar, A.~Mohammadi, L.~Akhtari and K.~Saaidi,
  ``Hamilton-Jacobi formalism to warm inflationary scenario'',
  Phys.\ Rev.\ D {\bf 95} (2017) no.2,  023501
  [arXiv:1708.01714 [gr-qc]].
  


\bibitem{Linde:1983gd}
  A.~D.~Linde,
  ``Chaotic Inflation'',
  Phys.\ Lett.\  {\bf 129B} (1983) 177.


\bibitem{Herrera:2018cgi}
  R.~Herrera,
  ``Reconstructing warm inflation'',
  [arXiv:1801.05138 [gr-qc]].





\end{thebibliography}
\end{document}